\documentclass[twocolumn, 9pt,a4paper]{article}

\usepackage[english]{babel}

\usepackage[a4paper]{geometry}

\usepackage{amsmath}
\usepackage{graphicx}
\usepackage[colorlinks=true, allcolors=blue]{hyperref}

\usepackage{listings}
\usepackage{xcolor}
\definecolor{codegreen}{rgb}{0,0.6,0}
\definecolor{codegray}{rgb}{0.5,0.5,0.5}
\definecolor{codepurple}{rgb}{0.58,0,0.82}
\definecolor{backcolour}{rgb}{0.95,0.95,0.92}

\lstdefinestyle{mystyle}{
    backgroundcolor=\color{backcolour},
    commentstyle=\color{codegreen},
    keywordstyle=\color{magenta},
    numberstyle=\tiny\color{codegray},
    stringstyle=\color{codepurple},
    basicstyle=\ttfamily\footnotesize,
    breakatwhitespace=false,
    breaklines=true,
    captionpos=b,
    keepspaces=true,
    numbers=left,
    numbersep=4pt,
    showspaces=false,
    showstringspaces=false,
    showtabs=false,
    tabsize=1,
    escapeinside={!!}
}

\usepackage{lineno}

\title{\textbf{pyMSER - An open-source library for automatic equilibration detection in molecular simulations}}
\author{Felipe Lopes Oliveira$^{a,b}$, Binquan Luan$^{c}$, \\Pierre Mothé Esteves$^{b}$, Mathias Steiner$^{a}$, Rodrigo Neumann Barros Ferreira$^{a,*}$}

\date{
	{\small $^{a}$IBM Research, Av. República do Chile, 330, CEP 20031-170, Rio de Janeiro, RJ, Brazil.\\
            $^{b}$Instituto de Química, Universidade Federal do Rio de Janeiro,  Av. Athos da Silveira Ramos, 149, CT A-622, Cid. Univ., Rio de Janeiro, RJ, 21941-909 - Brazil\\
            $^{c}$IBM Research, 1101 Kitchawan Rd, Yorktown Heights, NY 10598, United States\\
		*Corresponding author: rneumann@br.ibm.com}}

\begin{document}

\twocolumn[
    \begin{@twocolumnfalse}
    \maketitle
    \begin{abstract}

    Automated molecular simulations are used extensively for predicting material properties. Typically, these simulations exhibit two regimes: a dynamic equilibration part, followed by a steady state. For extracting observable properties, the simulations must first reach a steady state so that thermodynamic averages can be taken. However, as equilibration depends on simulation conditions, predicting the optimal number of simulation steps \textit{a priori} is impossible. Here, we demonstrate the application of the Marginal Standard Error Rule (MSER) for automatically identifying the optimal truncation point in Grand Canonical Monte Carlo (GCMC) simulations. This novel automatic procedure determines the point in which steady state is reached, ensuring that figures-of-merits are extracted in an objective, accurate, and reproducible fashion. In the case of GCMC simulations of gas adsorption in metal-organic frameworks, we find that this methodology reduces the computational cost by up to 90\%. As MSER statistics are independent of the simulation method that creates the data, this library is, in principle, applicable to any time series analysis in which equilibration truncation is required. The open-source Python implementation of our method, \texttt{pyMSER}, is publicly available for reuse and validation at \href{https://github.com/IBM/pymser}{https://github.com/IBM/pymser}.

    \vspace{25px}
    \end{abstract}

\end{@twocolumnfalse}]

\section{Introduction}

    Computer simulations have proven to be an extremely powerful tool to accelerate the study and discovery of new materials.\cite{ongari2020too} In particular, Grand Canonical Monte Carlo (GCMC) is one of the most used methods to perform molecular-based simulations of the adsorption of gases and other molecules into nanoporous materials such as zeolites, metal-organic frameworks (MOF), and covalent-organic frameworks (COF).\cite{watanabe2012accelerating, farmahini2018crystal, daglar2020recent, ren2022high}

    In GCMC simulations~\cite{dubbeldam2013inner} the system starts in a transient, \textit{out-of-equilibrium} condition, and random moves such as insertion, deletion, rotation, and translation are attempted on the adsorbed molecules. Such attempts can be accepted or rejected depending on the energy and temperature of the system, driving it towards an equilibrium state. Each attempted move, whether accepted or not, comprises a ``Monte Carlo step''. After a certain number of steps, the system reaches a stage where the properties of interest fluctuate around an equilibrium value.

    This creates a scenario in which the simulation can be split by a truncation point in two well-defined stages: \textit{i)} the \textit{``equilibration''} stage, which holds the initial transient configurations before the system achieves thermodynamic equilibrium and thus needs to be discarded, and \textit{ii)} the \textit{``production''} stage, that contains system configurations in thermodynamic equilibrium and is used to calculate the ensemble average of the properties of interest.\cite{frenkel2001understanding}

\begin{figure*}
        \centering
        \includegraphics[width=1\textwidth]{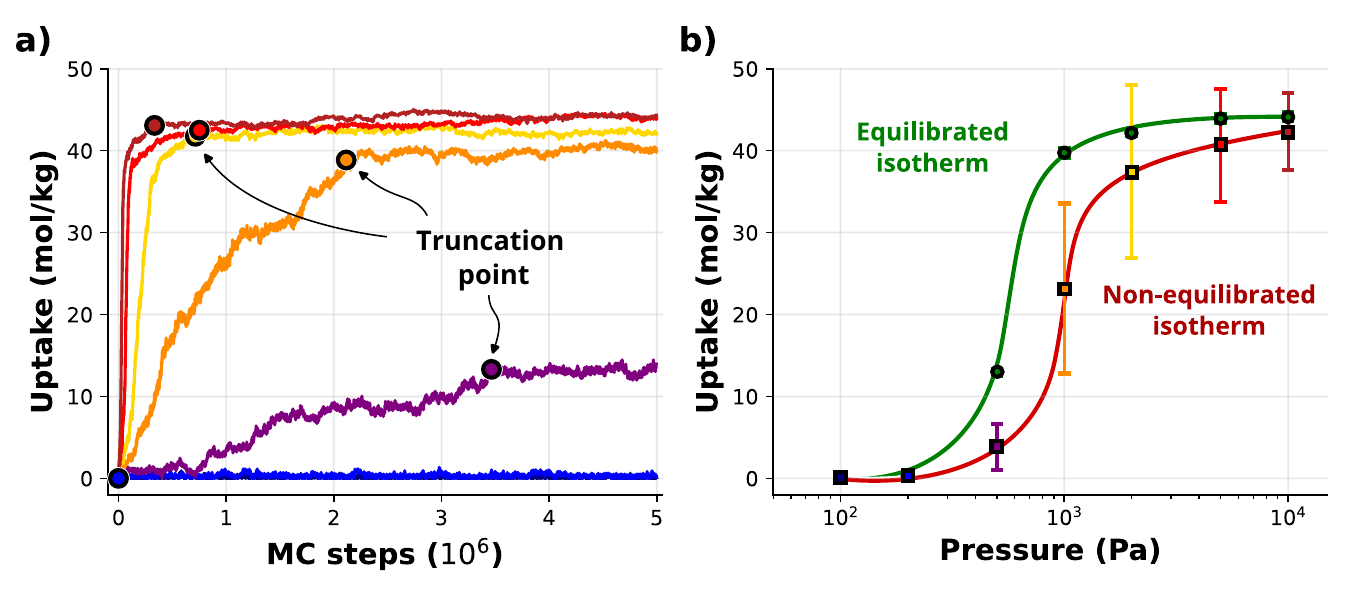}
        \caption{\textbf{Dependence of truncation points on simulation conditions.} \textbf{a)} Gas uptake as a function of number of MC steps. The black circles mark the truncation point, after which the production stage starts. \textbf{b)} Isotherm curves built using uptakes averaged over the entire time series (non-equilibrated) and over the production stage (equilibrated). Colored squares represent the uptakes averaged over the full time series without removing the initial transient data points. Green circles represent the uptakes obtained by averaging only over the production stage. Error bars from the properly averaged data lie within the symbol size.}
        \label{fig:gcmc_comparison}
    \end{figure*}

    While most GCMC simulation software allows users to define a fixed number of Monte Carlo (MC) steps for the equilibration and production stages, the optimal number of equilibration steps required to reach equilibrium cannot be predicted \textit{a priori}. The number of equilibration steps is sensitive to a variety of simulation conditions, including the adsorbed phase composition, temperature, pressure, and the frequency of each type of move, as shown in \autoref{fig:gcmc_comparison}(a). Every single pressure point in a simulated isotherm curve might require a different number of steps to achieve thermodynamic equilibrium. This makes the optimal number of MC steps in each stage impossible to anticipate, requiring a case-by-case analysis and the need for human intervention.

    If too few equilibration steps are used, the simulation only reaches the equilibrium after the truncation point, hence well into the production stage. This causes non-equilibrated states to be included in the ensemble average taken over the (improperly curated) production stage, which negatively affects the accuracy of the material property estimates so obtained. This situation is exemplified in \autoref{fig:gcmc_comparison}(b), where the isotherm (colored circles) generated without removing the equilibration stage underestimates the uptake results when compared to the well-equilibrated uptake values (green squares) calculated as the average over the production stage.

    On the other hand, when an excessively large number of equilibration steps is used to ensure that the system reaches the equilibrium before calculating the average, the computational workload is unnecessarily burdened by the poor choice of simulation parameters. This is particularly undesired in high-throughput computational studies in which millions of simulations are performed, and a few more steps in each simulation will have a huge impact on the overall computational cost and simulation time.

    Currently, the prevailing practice in the literature is to define a fixed number of equilibration steps, often ranging from 1,000 to 50,000~\cite{moghadam2018computer, avci2018high, ahmed2019exceptional, leperi2019development, li2019screening, ongari2019building, altintas2020role, do2005adsorption, pyzer2021accelerating, zhou2021modeling, chen2021computational, umeh2022eleven, han2023design, ercakir2024high}. Although for low pressures a smaller number of equilibration steps may be sufficient, one cannot know whether these arbitrarily chosen numbers of steps are sufficient to ensure accurate ensemble averages across all simulation conditions. This becomes especially critical for simulations where the number of molecules within the pore is very large, such as at higher pressures and/or lower temperatures. While some approaches have been proposed to determine the appropriate truncation point in molecular simulations, no single method has yet emerged as the widely accepted or standardized approach. 

    Gowers et al.~\cite{gowers2018automated} proposed an automated method to ensure that ensemble averages are calculated only over the production stage of a molecular simulation. This method involves first performing a very long simulation ($250 \times 10^6$ MC steps) and calculating the average and standard deviation of the second half of the data. Then, the moving average of the number of molecules in the system is calculated using a window of $2 \times 10^4$ MC steps over the first half of the data and compared to the average of the second half of the data. The truncation point is considered to be the point at which the moving average over the first half is within two standard deviations of the average over the second half. While this approach effectively distinguishes between the equilibration and production stages, it relies on the execution of a very long simulation making it prohibitive for high-throughput studies or simulations based on more computational-intensive approaches.

    Martin \textit{et al.}~\cite{yang2004free} proposed a method called \textit{Reverse Cumulative Averaging} (RCA), in which a property of interest is monitored in the reverse direction starting from the last frame of the time series and defines the ``border'' of the production step the point where the confidence limit in the normality of the observed value falls below 90\%. While this method could be applied to GCMC simulation data, it has several limitations. First, the truncation point is determined based on an arbitrary desired precision, which creates an inherent dependence of the truncation point on the chosen precision, making the methodology less robust. Second, the RCA method assumes that the desired precision can be achieved with the available simulation data, which may not always be possible. Third, the RCA method is sensitive to data noise, and as it proceeds in the reverse direction, it may underestimate the correct truncation point.

    The use of the multi-state Bennett acceptance ratio (MBAR) has been suggested for automated detection of the equilibrium stage in molecular dynamics simulations~\cite{chodera2016simple}. Despite its potential benefits, the high computational cost of the MBAR approach renders it unsuitable for high-throughput screening studies. Furthermore, it has been observed that even in straightforward simulation data, this method falls short of accurately identifying the correct production stage, as will be illustrated later.

    Here we propose the use of the Marginal Standard Error Rule to automatically detect the optimal truncation point for the production stage. This approach is simple, fast, robust, and reproducible, and it allows for the exclusion of the equilibration stage and the identification of the maximum achievable production stage for a given total number of simulation steps. We have implemented this method as an open-source Python package called \texttt{pyMSER}, and demonstrate its effectiveness on various examples of typical GCMC simulation data, particularly for calculating adsorption uptake and enthalpy of adsorption. Additionally, it can be used to develop on-the-fly tests that can automatically detect the truncation point and continue the simulation until a desired number of production steps is obtained.

\section{Methodology}
    \subsection{GCMC Simulations}

    The Grand Canonical Monte Carlo simulations were performed using RASPA (version \texttt{v2.0.45}).\cite{dubbeldam2013inner, dubbeldam2016raspa} Interaction energies between non-bonded atoms were computed via a combination of Lennard-Jones (LJ) and Coulomb potentials

    \begin{equation}
        U_{ij}(r_{ij}) = 4\varepsilon_{ij} \left [  \left( \frac{\sigma_{ij}}{r_{ij}} \right)^{12}  - \left( \frac{\sigma_{ij}}{r_{ij}} \right)^6\right] + \frac{1}{4\pi \epsilon_0} \frac{q_i q_j}{r_{ij}}
    \end{equation}
    
    where $i$ and $j$ are the atom indexes and $r_{ij}$ is their inter-atomic distance. $\varepsilon_{ij}$ and $\sigma_{ij}$ are the LJ potential well depth and zero-energy radius, respectively, $q_i$ and $q_j$ are the partial charges of atoms $i$ and $j$\textcolor{black}{, and $\epsilon_0$ is the electric constant}. LJ parameters for framework atoms were taken from the Universal Force Field (UFF) \textcolor{black}{with the values provided in Supporting Information}~\cite{rappe1992uff}. The partial charges were calculated using the EQeq method~\cite{wilmer2012extended}. The parameters for the adsorbed molecules were taken from the TraPPE~\cite{potoff2001vapor} force field. The LJ parameters between atoms of different types were calculated using the Lorentz-Berthelot mixing rules.
    
    \begin{equation}
        \varepsilon_{ij} = \sqrt{\varepsilon_{ii}\varepsilon_{jj}}, \qquad \sigma_{ij} = \frac{\sigma_{ii} + \sigma_{jj}}{2}
    \end{equation}

    A cut-off radius of 12.8 {\AA} was used for LJ and charge-charge short-range interactions and the Ewald sum technique with a relative precision of $10^{-6}$ was used to compute the long-range electrostatic interactions. The Lennard-Jones potential was shifted to zero at the cut-off radius. The framework atoms were held fixed at their initial positions, as described by their respective Crystallographic Information Files. Fugacities used by the RASPA code to impose equilibrium between the system and the external ideal gas reservoir at each pressure were calculated using the Peng-Robinson equation of state.\cite{peng1976industrial}

    All simulations were performed with Monte Carlo step sampling such that swap (insertions or deletions), translations, rotations, and re-insertions were performed with probabilities of 0.5, 0.3, 0.1, and 0.1 respectively. The results that have been presented are analyzed with respect to the number of MC steps. The MC steps were converted from RASPA cycles. Each RASPA cycle contains $max(20, N)$ MC cycles where $N$ is the number of atoms present in the simulation box.

    \subsection{The Marginal Standard Error Rule}

    The Marginal Standard Error Rule (MSER), originally called the Marginal Confidence Rule, comprises a family of initial transient detection algorithms based on the minimization of the marginal confidence interval over the truncated sample mean of the data.\cite{white1997effective} The essential idea behind the MSER approach is the assumption that the initial values of a given simulated observable $Y$, \textit{e.g.} the number of molecules adsorbed on a porous material, are far from the steady-state mean. Thus, removing the initial $k$ values of the series increases the accuracy of the mean by removing the biased initial conditions of the simulation. On the other hand, removing more values than the initial non-equilibrated ones leads to a reduction of the sample size, which will subsequently decrease the confidence in the average value.

\begin{figure}[!ht]
        \centering
        \includegraphics[width=.5\textwidth]{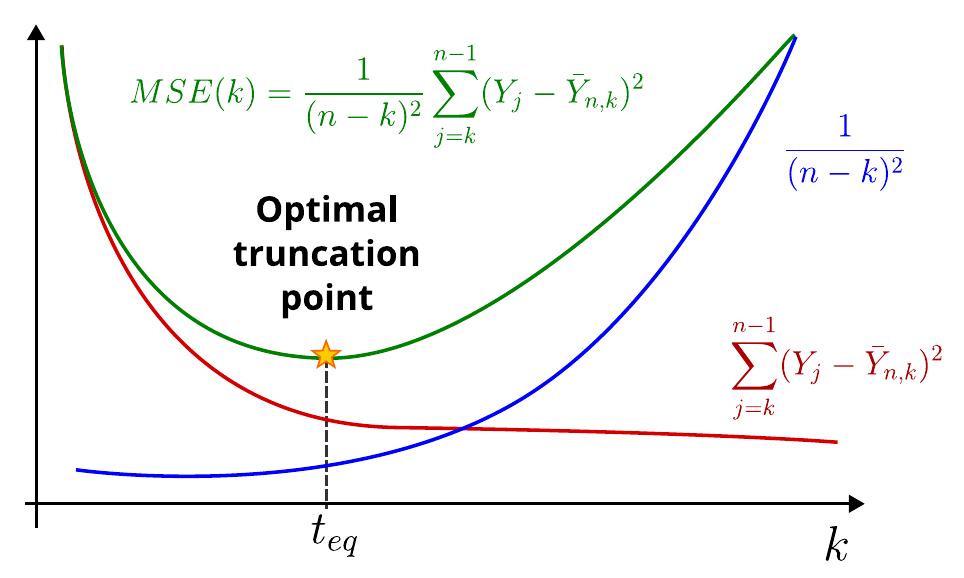}
        \caption{\textbf{Detection of the optimal truncation point.} The balance between improved accuracy (red curve), by removing transient data, and improved confidence (blue curve), by increasing the sample size, leads to the determination of the optimal truncation point $t_{\text{eq}}$ that minimises the $\text{MSE}(k)$ curve (green curve).}
        \label{fig:mse}
    \end{figure}

    The balance between improved accuracy, as a consequence of bias reduction, and decreased confidence, as a consequence of sample size reduction, is obtained at an optimal truncation point that is the one that minimizes marginal standard error over the truncated sample mean, as stated in \autoref{t_eq} and illustrated in \autoref{fig:mse}. Formally, this algorithm proposes that the optimal truncation point, here named $t_{\text{eq}}$, for a finite data series with size $n$ \{$Y_j$: $j=0, 1, 2, ..., n$\} is the point that minimizes the Marginal Standard Error after deleting the first $k$ points of the data, here represented as $\text{MSE}(k)$:

    \begin{equation}
        \label{t_eq}
        t_{\text{eq}} =  \underset{1 \leq k \leq n-2}{\mathrm{arg\,min}} \, \text{MSE}(k)
    \end{equation}
        
    where the $\text{MSE}(k)$ is calculated as

    \begin{equation}
        \label{mse_eq}
        \text{MSE}(k) = \frac{1}{(n - k)^2} \sum_{i=k}^{n-1}(Y_i - \bar{Y}_{n,k})^2
    \end{equation}

    \textcolor{black}{where $n$ represents the number of points in the original series, $Y_i$ denotes the property value at the $i$-th point, and $\bar{Y}_{n,k}$ the average of the remaining data points.}

    The main assumption necessary to ensure the validity of this method is the existence of a unique steady-state within the data series regardless of the truncation point selected.

    There are two MSER variants, in addition to the original version, designed to present better performance on specific use-cases. The first variant, MSER-LLM, aims to preserve as much of the original data series as possible. This is achieved by selecting the \textbf{L}eft-most \textbf{L}ocal \textbf{M}inimum of the MSER curve as the optimal truncation point instead of the global minimum, as in the original MSER version. This strategy generates a good balance between sample size increase and bias elimination, removing less data than the original method with a minimal impact on the final average accuracy, as will be shown later.

    The second MSER variant enhances the execution performance by batch-averaging the number of evaluated points, drastically reducing the execution time.\cite{pasupathy2010initial} This can be obtained by creating non-overlapping batches of size $m$, taking the average of each batch, and calculating the respective MSER on the series of batch averages.\cite{white2000comparison} The only drawback is the loss of precision over the location of $t_{\text{eq}}$, which then can only be done to within $m$ steps. This approach can be applied to both MSER and MSER-LLM, thus generating the variants MSER-\textit{m} and MSER-LLM-\textit{m}. The partitioning of the initial data set into batches can be very useful to increase the stability of the method since it reduces the noise present in the data and also reduces the already relatively low computational cost. This facilitates its use in long simulations, high-throughput screening, and on-the-fly equilibrium verification.

    \subsection{Statistical uncertainty}

    Calculating the accuracy of an estimated property is critical for deriving reliable conclusions, especially when comparing the simulations to experimental values. A common choice to represent the uncertainty of an average property obtained from a GCMC simulated time series is the Standard Deviation (SD). Since the number of molecules obeys a Gaussian distribution around a mean value during the production stage, it seems natural to convey the uncertainty as the SD of that distribution. However, as each new MC step is obtained by applying a tiny random update over the previous step, the values in this Markov Chain are strongly correlated, and thus the SD can overestimate the real uncertainty of the simulation.\cite{flyvbjerg1989error}

    To reduce the impact of correlation on the uncertainty calculation, we propose the use of the uncorrelated Standard Deviation (uSD). The uSD is obtained by accounting for the ``autocorrelation time'' $\tau$ of a given property $Y$. The autocorrelation time is obtained by calculating the normalized autocorrelation function (ACF) and then fitting an exponential decay using a least-squares minimization technique.

    \begin{equation}
        \text{ACF}(t) = \frac {\langle Y(t_0) Y(t_0 + t)\rangle - \langle Y \rangle ^2} {\langle Y^2 \rangle - \langle Y \rangle ^2} \approx e^{-t/\tau}
        \label{autocorrelation}
    \end{equation}

    \textcolor{black}{where $\tau$ represents the ``autocorrelation time'' and $Y(x)$ denotes the property value at the $x$-th point.}

    The number of statistically uncorrelated samples is given by $N_{\text{u}} = N_{\text{prod}} / 2 \tau$, where $N_{\text{prod}}$ represents the total number of steps in the production stage over which the ensemble averages are taken, and $2 \tau$ is an estimate of the number of steps required to obtain a new statistically uncorrelated sample.\cite{janke2002statistical} The production time series is divided into $N_{\text{u}}$ blocks containing $2 \tau$ data points and the block averages are used to build a (new) uncorrelated time series. The uSD is therefore calculated as the SD of the uncorrelated time series.

\begin{figure}[!ht]
        \centering
        \includegraphics[width=1\textwidth]{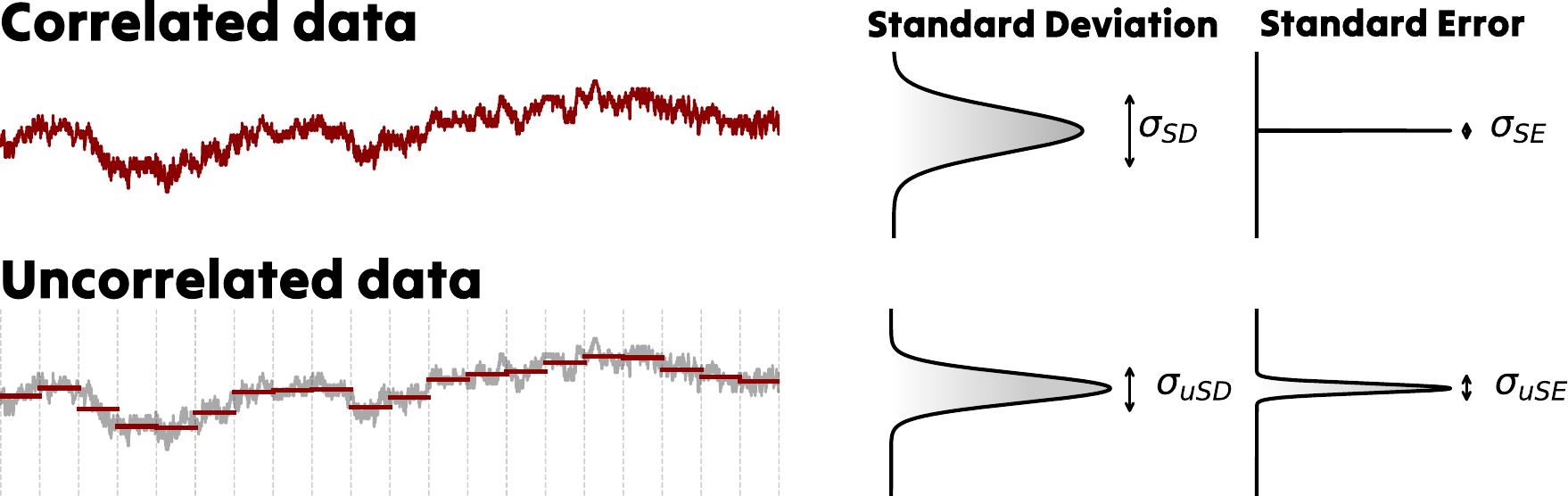}
        \caption{\textbf{Correlated and uncorrelated uncertainty metrics.} The upper plots show the correlated data (left) and the standard deviation/error histogram (right), respectively, for a typical simulated time series. The bottom plots show the uncorrelated average data (left) and the respective standard deviation and standard error metrics (right).}
        \label{fig:uncertainty}
    \end{figure}

    The same process can be carried out for the Standard Error of the mean (SE), defined as $SD/\sqrt{N_{\text{prod}}}$, thus generating the uncorrelated Standard Error of the mean (uSE). \autoref{fig:uncertainty} illustrates the process of creating uncorrelated batches and its impact on the uncertainty metrics. \textcolor{black}{We note that the selection of the uncertainty metric employed in the calculation does not impact the estimation of the truncation point or the average value.}
    
\section{Results and discussion}
    
    \paragraph{Effects of MSER Variants}

\begin{figure}[!ht]
        \centering
        \includegraphics[width=.5\textwidth]{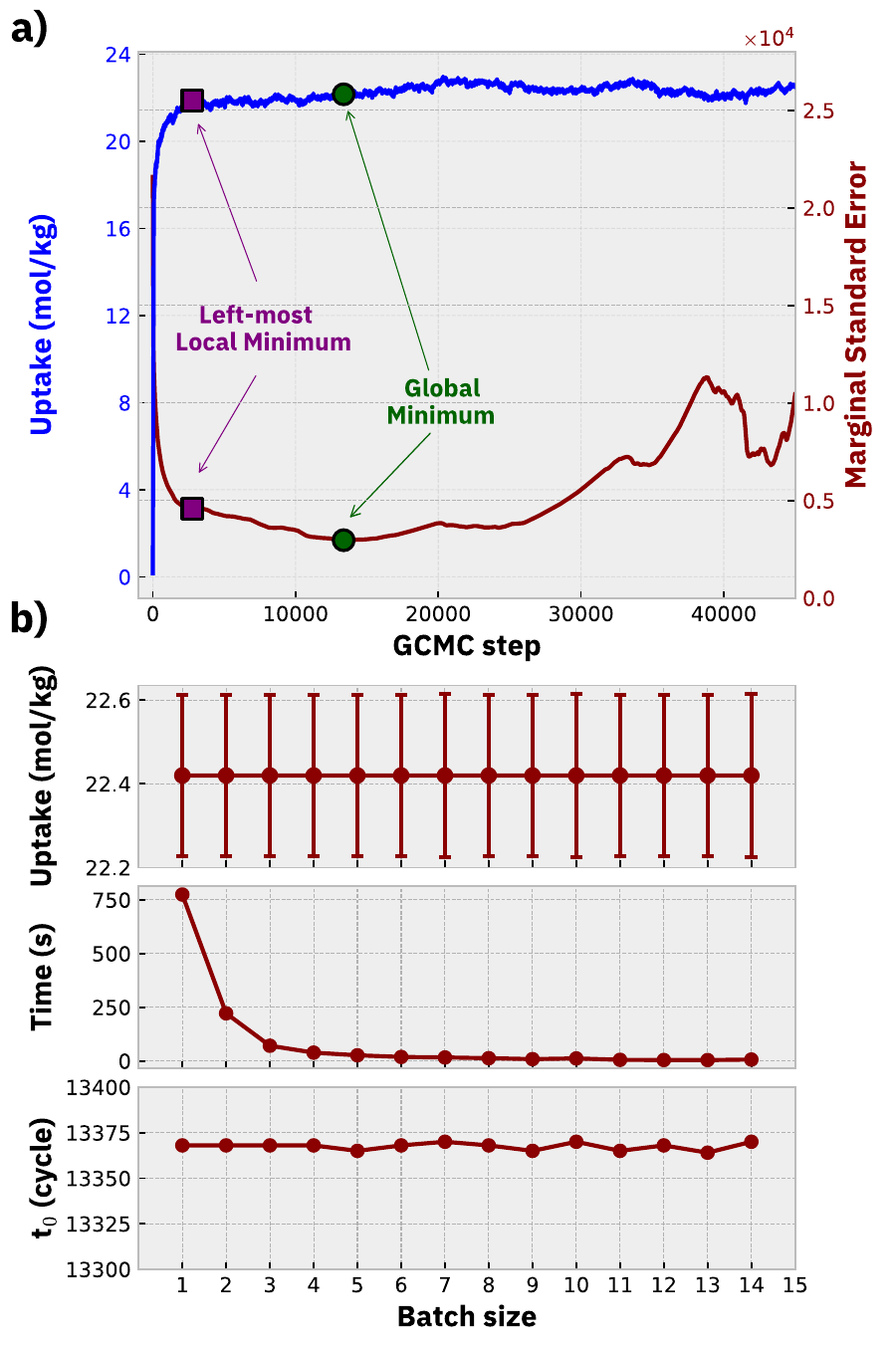}
        \caption{\textbf{Effects of using MSER-LLM and MSER-$m$ variants on simulated CO\textsubscript{2} adsorption data}. \textbf{a)} The dark blue (red) curve represents CO\textsubscript{2} uptakes (MSE) at each MC step. The purple square marks the left-most local minimum and the green circle marks the global minimum of the MSE curve. \textbf{b)} Resulting average uptake, execution time and truncation point after running the MSER-$m$ variant, as a function of batch size $m$.}
        \label{uptake_example}
    \end{figure}

    We applied each MSER variant to a reference dataset to assess their impact on the observable properties. In \autoref{uptake_example}(a) we demonstrate how MSER variants can be applied to GCMC simulations, exemplified by evaluating the adsorption of CO\textsubscript{2} on the metal-organic framework HKUST-1. From a simple visual inspection, the uptake value reaches a steady-state mean after approximately 2000 MC steps. In fact, the MSER curve has its left-most local minimum at step number $\sim 2700$, whereas the global minimum is at step number $\sim 13300$. Despite the significant difference in the truncation point between the original MSER and the MSER-LLM variants, the average uptakes over the production region ($t \ge t_{\text{eq}}$) are compatible, being $\left( 22.4 \pm 0.2 \right)\,\text{mol/kg}$ and $\left( 22.3 \pm 0.3 \right)\,\text{mol/kg}$, respectively. Therefore, even though the MSER-LLM variant removes less transient data, it has little impact on the average uptake value.

    The utilization of non-overlapping batches in the MSER-\textit{m} variant does not generate any significant change in the average uptake, uncertainty estimate, or in the position of the truncation point, as shown in~\autoref{uptake_example}(b). Nevertheless, the execution can be 10 to 25 times faster depending on the $m$ value chosen, being therefore strongly beneficial to high-throughput studies. We recommend the use of $m \approx 0.0005 N$, where $N$ is the total number of data points in the time series. The use of larger $m$ is not recommended without the LLM version, as it can generate instability in the results without providing a relevant reduction in execution time. For mode details see Supporting Information.

    \paragraph{Application to GCMC Data}

\begin{figure*}[!ht]
            \centering
            \includegraphics[width=1\textwidth]{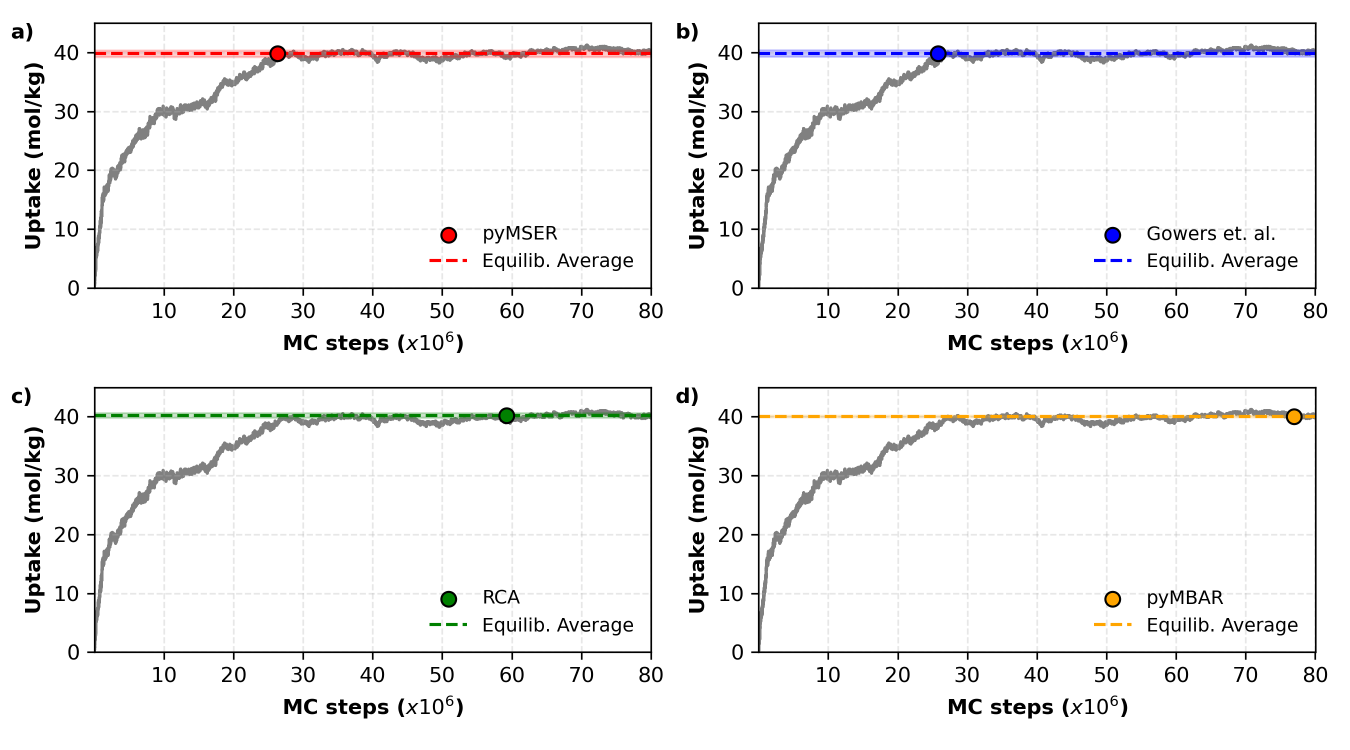}
            \caption{\textbf{Truncation point detection with different approaches.} Both pyMSER (a) and the method suggested by Gowers \textit{et al.} (b) identify the truncation point around the same value ($\sim 26 \cdot 10^6$ steps) while the RCA (c) and pyMBAR (d) detects a very late truncation point for the production stage. \textcolor{black}{These methodologies were applied to GCMC simulation data for H\textsubscript{2}O adsorption in MOF HKUST-1 at 273K.}} 
            \label{diff_approaches}
        \end{figure*}

        To explore the potential advantages of using pyMSER over other approaches,~\autoref{diff_approaches} compares four methodologies for obtaining equilibrated averages -- namely, pyMSER, Gowers \textit{et al.}, RCA, and pyMBAR -- applied to the GCMC simulation of H\textsubscript{2}O adsorption on the HKUST-1 material. The results show that both pyMSER and the methodology recommended by Gowers \textit{et al.} produced similar results, identifying the beginning of the production region after $\sim 26 \times 10^6$ steps. Despite the equivalent results, it is important to point out that the method by Gowers \textit{et al.} requires a large number of simulation steps (e.g. $250 \times 10^6$) to ensure that the entire second half of the simulation is on the equilibrated region. In contrast, MSER only requires that an equilibrated region exists within the adsorption time series. Consequently, pyMSER can produce equivalent results with significantly less simulation cost. 
        
        When the same analysis is repeated with a shortened version of the time series, as in Figure S5, Gowers \textit{et al.}, RCA, and pyMBAR methodologies have produced varying outcomes. pyMSER, on the other hand, has generated results identical to those from the full time series. This highlights the reliability and consistency of pyMSER in producing accurate results.
        
        In contrast, the RCA and pyMBAR methods have identified the truncation point to be present only in the latter stages of the simulations, specifically around $60 \times 10^6$ and $77 \times 10^6$ steps, respectively, as shown in \autoref{methods}. Although the mean values obtained within the equilibrated region detected by these methods are numerically similar to those obtained using pyMSER, it is apparent that the truncation points detected by these methods are not accurate or optimal for this simulation. This leads to a strong underestimation of the uncertainty obtained with these methods and prevents its use as a tool for evaluation of equilibration in GCMC simulations. The detection of a truncation point on the late portion of the time series is likely related to the fact that both methodologies assess the time series in reverse direction and does not make a global assessment of the data. This makes both methodologies more susceptible to noise and the normal oscillations present in GCMC simulations. More details about the application of these methods on other simulation scenarios are presented as Supporting Information.

        \begin{table}[ht!]
            \centering
            \begin{tabular}{l|c|c|c}
                \hline\hline
                \textbf{method} & \textbf{time (s)} & \textbf{t\textsubscript{0}} ($\times$10\textsuperscript{6}) & \textbf{uptake (mol/kg)} \\ \hline
                Gowers          & 0.05   & 25.7 & 39.85 $\pm$ 0.56 \\
                pyMSER          & 0.16   & 26.3 & 39.80 $\pm$ 0.56 \\
                RCA             & 0.51   & 59.2 & 40.24 $\pm$ 0.24 \\
                pyMBAR          & 78.93  & 76.9 & 40.07 $\pm$ 0.14 \\
                \hline \hline
            \end{tabular}
            \caption{Comparison of execution times, truncation point, uptake averages and uncertainties obtained with different equilibration methods from the same time series data.}
            \label{methods}
    \end{table}

        In the case of RCA, a relative precision value of 0.015 was employed. Experiments conducted by varying this value revealed a significant dependence of the truncation point on the selection of this parameter, thereby emphasizing the necessity for a careful examination of individual cases to obtain dependable outcomes. Regarding pyMBAR, experiments that involved altering the number of samples employed to sparsify data or other adsorption data did not lead to any notable improvements in results.

    \paragraph{Open-source Python Implementation}

    The \texttt{pyMSER} package is a pure Python implementation of the MSER algorithm, that can be used for the automatic identification of the truncation point in any time series. This package can apply the original MSER algorithm as well as the MSER-LLM and MSER-\textit{m} variants to find the truncation point in any type of data, provided that there is a segment of the data where the instantaneous values fluctuate around a single steady state with a well-defined mean value. Here, the application of pyMSER to adsorption data from GCMC simulations was specifically explored. Nevertheless, these algorithms can be applied in principle to any type of data regardless of its nature.

    pyMSER makes use of the \texttt{numpy}\cite{harris2020array}, \texttt{scipy}\cite{2020SciPy-NMeth} and \texttt{torch}\cite{paszke2019pytorch} libraries to deal with the numerical array calculations and the \texttt{statsmodels}\cite{seabold2010statsmodels} library to apply the Augmented Dickey-Fuller test. One can easily install pyMSER from PyPI (i.e., \texttt{pip install pymser}), and the source code is also available from the GitHub repository (\url{https://github.com/IBM/pymser}) under a \texttt{BSD-3-Clause} license to allow for further development.

    The automated detection of the optimal truncation point with MSER is implemented inside the \texttt{equilibrate()} function. This function can be applied to any time series data using code similar to the one presented in~\autoref{MSER}. The \texttt{LLM} keyword is a Boolean variable that controls the activation of the MSER-LLM variant, the \texttt{batch\_size} keyword allows the use of non-overlapping batches (MSER-\textit{m} and MSER-LLM-\textit{m} variants) of integer size $m$ to reduce the execution time, and the \texttt{uncertainty} keyword allows the choice of the metric for estimating the uncertainty of the average values reported. Although here we argued for the use of the uncorrelated Standard Deviation (uSD), other options such as uncorrelated Standard Error (uSE), Standard Deviation (SD), and Standard Error (SE) are also available. The execution usually takes a few seconds and the results are stored in the form of a dictionary, allowing a direct evaluation of the results and a simple integration into complex simulation workflows.

    We employed the Augmented Dickey-Fuller (ADF) test~\cite{greene2003econometric} to independently verify the stationary nature of the production stage identified by pyMSER. The ADF test is designed to assess whether a given data set is non-stationary by comparing it to a critical value, in order to evaluate if the null hypothesis -- that the data is non-stationary -- can be rejected. If the null hypothesis is rejected, it implies that the time series is stationary. A non-stationary time series has a mean, variance, and autocovariance that change over time. The ADF test is performed when the keyword \texttt{ADF\_test=True}.

    \textcolor{black}{An alternative to the ADF test is the Gelman-Rubin method~\cite{anstine2020screening}, which can also be used to verify the equilibration of the production stage. We included an in-depth analysis of the Gelman-Rubin method to the Supporting Information section.}

    \begin{figure}[!ht]
        \centering
        \begin{lstlisting}[language=Python]
import pymser
import pandas as pd

# Load the .csv file
df = pd.read_csv('example_data/Cu-BTT_500165.0_198.000000.csv')

results = pymser.equilibrate(df['mol/kg'], 
                              LLM=False, 
                              batch_size=1, 
                              ADF_test=True, 
                              uncertainty='uSD', 
                              print_results=True)
    \end{lstlisting}
        \caption{\textbf{pyMSER usage example.} The \texttt{equilibrate()} function applies the MSER method to the adsorption data and generates a small report with the results. This example uses the time series data provided in the repository.}
        \label{MSER}
    \end{figure}

    \paragraph{Use in High-throughput Simulation Workflows}

        The main use of pyMSER may lie in the post-processing of simulation data, in which only after the simulation finishes, the equilibration stage is removed and the remaining data -- the production stage -- is used to calculate the desired average value. This approach has been employed in the creation of the CRAFTED database,~\cite{oliveira2023crafted, lopes_oliveira_felipe_2023_7689919} that contains 1,270,152 independent GCMC simulations in which the equilibrated average and standard deviation was obtained using pyMSER. 

        Another potential use of pyMSER is to be integrated into simulation workflows, where the simulation is monitored at regular intervals to detect the start of the production stage. Then, a given number of production steps can be performed in order to obtain a well-equilibrated property. \textcolor{black}{An example of this on-the-fly analysis workflow is implemented in the GitHub repository provided in the Data and Software Availability section.}
        
        \begin{figure*}[!ht]
            \centering
            \includegraphics[width=.95\textwidth]{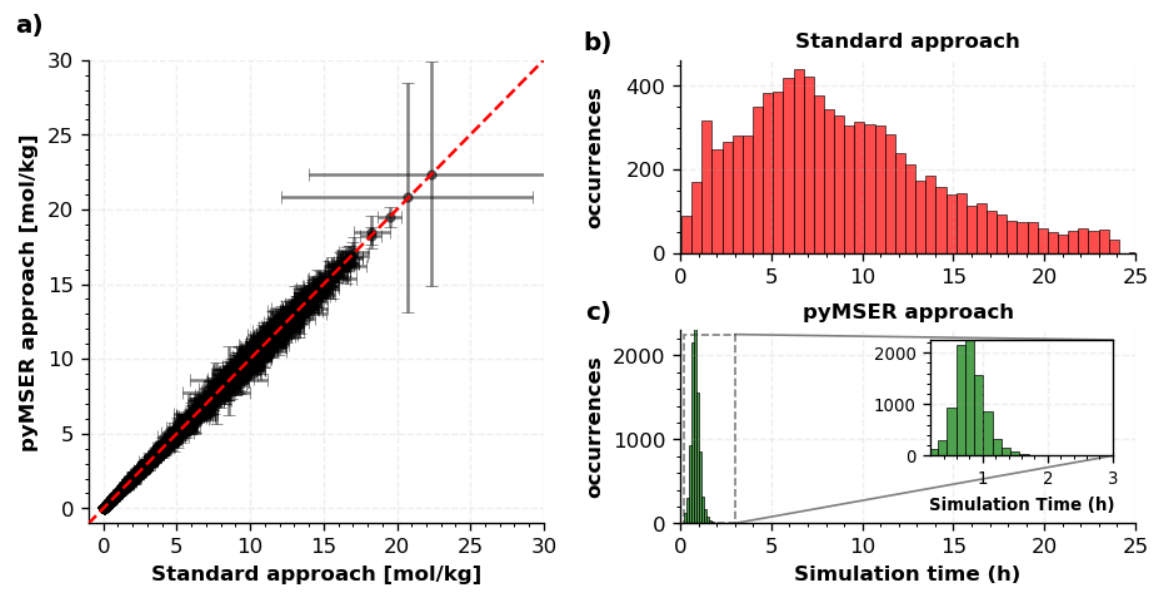}
            \caption{\textbf{Comparison of approaches for simulating CO\textsubscript{2} uptake on MOFs from the CoRE-MOF-2019 database.} \textbf{a)} Comparison between the CO\textsubscript{2} uptake obtained with two differente approaches. The on-the-fly MSER approach achieves the same results as the standard approach does, albeit with a tiny fraction of the computational cost due to running drastically less MC steps. \textbf{b-c)} Histograms comparing execution time using the standard and MSER approaches, demonstrating the significant reduction in total execution time achievable with the latter.}
            \label{fig:on_the_fly}
        \end{figure*}
        
        To evaluate the potential impact of \texttt{pyMSER} compared to the standard GCMC approach widely used in high-throughput simulations, let us consider the following scenario. A researcher wants to evaluate the carbon capture performance of MOF structures selected from the CoRE-MOF-2019 database~\cite{chung2019advances} by simulating the adsorption of CO\textsubscript{2} at 1 bar and 298 K. The researcher has two options: (i) use the standard method from the literature, which involves running a fixed number of MC equilibration cycles followed by a fixed number of MC production cycles, or (ii) develop a straightforward code that assesses, at specified intervals (e.g., every 1000 cycles), whether the simulation has attained equilibrium, and then runs a fixed number of MC production cycles.

        \textcolor{black}{We take, as the ``standard'' approach, one that finds ample usage in the literature.}~\cite{do2005adsorption, pyzer2021accelerating, zhou2021modeling, chen2021computational, umeh2022eleven, han2023design, ercakir2024high} \autoref{fig:on_the_fly}(a) illustrates the uptakes obtained using these two approaches: (i) the standard approach represented on the \textit{x}-axis\textcolor{black}{, with 50,000 equilibration cycles and 50,000 production cycles}, and (ii) the pyMSER approach depicted on the \textit{y}-axis\textcolor{black}{, with 10,000 production cycles simulated after reaching equilibration.} Additionally, \autoref{fig:on_the_fly}(b) present histograms comparing the execution times of these approaches, highlighting the substantial reduction achievable with the pyMSER method. \textcolor{black}{On average, the execution time using pyMSER is $\frac{1}{10}$ that of the standard approach, and it can be up to $\frac{1}{30}$ in some cases.} In this analysis, we discarded CoRE-MOF-2019 materials whose GCMC simulation took more than 24 hours due to limitations in the job submission queue system. The results demonstrate agreement within the standard deviation, indicating that despite its lower computational cost, the pyMSER approach yields consistent results comparable to well-established, albeit more computationally intensive, methodologies.

    \paragraph{Enthalpy of Adsorption}

        In addition to the uptake of adsorbed molecules on porous materials, another property that can be strongly impacted by using non-equilibrated data is the enthalpy of adsorption~\cite{snurr1993prediction}. The enthalpy of adsorption is a piece of important thermodynamic information about the adsorption process and can be obtained with minor computational cost from the existing simulation data on the energy and particle fluctuations, as both the number of molecules and total energies are already being calculated at every step.

        The enthalpy of adsorption $\Delta H_{\text{ads}}$ (or the heat of adsorption $q = -\Delta H_{\text{ads}}$) can be calculated~\cite{vlugt2008computing, torres2017behavior} as

        \begin{equation}
            \Delta H_{\text{ads}} = \left( \frac{\partial U}{\partial N}\right)_{V,T} - \langle U_g \rangle - \langle U_h \rangle - RT
        \end{equation}

        where $U$ represents the total energy of the system (framework and adsorbed molecules), $\langle U_g \rangle$ is the average energy of a single molecule in the gas phase, $\langle U_h \rangle$ is the average energy of the framework (host) system, $R$ is the universal gas constant and $T$ the temperature of the simulation.

        The energy/particle fluctuations in the grand-canonical ensemble can be used to express the absolute differential energy ($\partial U/\partial N$) as a function of the averages of the total energy and number of molecules \cite{nicholson1982computer, myers2002thermodynamics, poursaeidesfahani2016computation} for a single component system as

        \begin{equation}
            \left( \frac{\partial U}{\partial N}\right)_{V,T} = \frac{\left( \frac{\partial U}{\partial \mu}\right)_{V,T}}{\left( \frac{\partial N}{\partial \mu}\right)_{V,T}} = \frac{\langle U \cdot N\rangle_\mu - \langle U \rangle_\mu \langle N\rangle_\mu}{\langle N^2\rangle_\mu - \langle N\rangle^2_\mu }
        \end{equation}

        where $\langle \dots \rangle_\mu$ represents the averages in the grand-canonical ensemble ($\mu$ is the chemical potential), $N$ is the number of adsorbates on the simulation box, and $U$ is the total potential energy.

        In simulations where both the framework structure and the adsorbed molecules have rigid bonds, the total internal energy of the host and the isolated molecule, $U_h = U_g = 0$. Therefore the enthalpy of adsorption can be calculated as  

        \begin{equation}
            \Delta H = \frac{\langle U \cdot N\rangle_\mu - \langle U \rangle_\mu \langle N\rangle_\mu}{\langle N^2\rangle_\mu - \langle N\rangle^2_\mu } - RT
        \end{equation}

        This approximation is widely used in GCMC simulations and it has been shown that for non-flexible frameworks it has only a small influence on the adsorption at low loading.\cite{vlugt2002influence} 

        \textcolor{black}{In general, the uptake and total energy of a single-component gas adsorption simulation do not necessarily reach equilibrium at exactly the same time. Therefore, pyMSER analyzes both time series independently and uses the largest detected $t_{\text{eq}}$ as the truncation point for selecting the region to calculate the adsorption enthalpy. Figure S3 illustrates this behavior.}

        It is noteworthy that the addition of non-equilibrated data can lead to a severe overestimation of the adsorption enthalpy calculated using the energy/particle fluctuations approach, as depicted in Figure S4. It is noticeable that not only is there a greater standard deviation among all calculated points when transient initial data is not excluded, but also significant disparities in equilibrium enthalpy of up to 10 kcal/mol. This underscores the critical importance of a meticulous data analysis process to ensure the attainment of reliable and reproducible results.

    \paragraph{Further Considerations}
    
        In simulations of multi-component gas adsorption, each component may have a unique truncation point. However, the total uptake truncation point equals that of the component with the latest truncation point. Thus, even in cases with multiple components, a single evaluation of the total uptake is sufficient to determine when the simulation has reached the production stage.

        Finally, although there is no prior requirement on the data except that the system is ergodic, \textit{i.e.} the simulation proceeds towards a unique statistical equilibrium state, it is important to point out that a reasonable amount of data at the production stage must exist in order to obtain reliable results. In this regard, the use of a reduced number of MC steps to study adsorption processes on porous materials must be done carefully. Although numerically correct results can be obtained with a reduced number of MC steps, a good sample of the configuration space is needed in order to make accurate estimates of ensemble averages and free energies.

        In this context, pyMSER is an ideal tool for automatically and reliably detecting the optimal truncation point in each simulation, without the need for human intervention or the creation of complex heuristics. pyMSER can also be used as an on-the-fly testing tool in automated workflows, allowing for an objective evaluation of whether a given simulation has reached a certain convergence criterion defined by the user and can be stopped. Possible criteria would be achieving a certain number of data points in the production region or a certain signal-to-noise ratio between the observable value and its uncertainty. \textcolor{black}{We refer readers to the GitHub repository where they can find an example workflow depicting on-the-fly pyMSER analysis, which can serve as a starting point for the development of more complex workflows.}

        Additionally, the MSER statistics do not depend on the nature of the data and are method-agnostic, i.e. independent of the simulation software that provides the data, therefore it can be applied to any time series data, whenever removing initial transient values is necessary. \textcolor{black}{While extensive testing demonstrates pyMSER's robustness, its use beyond the scope explored here requires independent evaluation by the user. For this purpose, we recommend the application of the Augmented Dickey-Fuller test, as implemented in pyMSER, to evaluate the equilibration of the production stage.}

\section{Conclusions}

    This work explores the use of the Marginal Standard Error Rule (MSER) as a method for automatically detecting the start of the production stage in molecular simulations and presents its implementation as an open-source Python package called \texttt{pyMSER}. This methodology simplifies data post-processing, reduces computational cost, and eliminates the need for human intervention in high-throughput simulation workflows. \texttt{pyMSER} provides an objective way to determine the accuracy of the simulation and also increases the reproducibility and reliability of the results. The open-source Python implementation is available under a \texttt{BSD-3-Clause} license and is provided online, along with several examples, at \url{https://github.com/IBM/pymser}.    
    
\section*{Acknowledgments}
    FLO and PME acknowledge financial support from CAPES (Project 001), CNPq, and FAPERJ. 

\section*{Contributions}
    F.L.O. conceived the methodology, developed the Python code, executed the simulations, compiled and analyzed the data, and wrote the manuscript. R.N.B.F. conceived the methodology, analyzed the data and wrote the manuscript. B.L. conceived the methodology and wrote the manuscript. P.M.E. wrote the manuscript. M.S. wrote the manuscript. All authors contributed to the discussion, and gave approval to the final version of the manuscript.

\section*{Conflicts of interest}
    There are no conflicts of interest to declare.

\section*{Data and Software Availability}
    All the source codes and data presented in this manuscript are available under the \texttt{BSD-3-Clause} license at \href{https://github.com/IBM/pymser}{https://github.com/IBM/pymser}.

\bibliographystyle{unsrt}
\bibliography{sample}

\newpage
\onecolumn

\setcounter{section}{0}
\setcounter{figure}{0}
\setcounter{table}{0}
\renewcommand{\thefigure}{S\arabic{figure}}
\renewcommand{\thetable}{S\arabic{table}}
\renewcommand{\thepage}{S-\arabic{page}}
\renewcommand{\thesection}{S\arabic{section}}

\begin{center}
\title{{\Huge Supplementary Information}}
\end{center}

 \section*{Example of pyMSER use}
    
    There are two steps in the example below. First, we read the CSV file containing the example data available from \url{https://github.com/IBM/pymser/tree/main/example_data}. Then, we use pyMSER to find the truncation point of the time series by using the \texttt{equilibrate()} function.
    
\begin{lstlisting}[language=Python]
>>> import pymser
>>> import pandas as pd

>>> # Load the .csv file
>>> df = pd.read_csv('example_data/Cu-BTT_500165.0_198.000000.csv')

>>> results = pymser.equilibrate(df['mol/kg'], 
>>>                              LLM=False, 
>>>                              batch_size=1, 
>>>                              ADF_test=True, 
>>>                              uncertainty='uSD', 
>>>                              print_results=True)

                             pyMSER Equilibration Results
 ==============================================================================
 Start of equilibrated data:          13368 of 48613
 Total equilibrated steps:            35245  (72.50%)
 Equilibrated:                        Yes
 Average over equilibrated data:      22.4197 +- 0.1905
 Number of uncorrelated samples:      22.3
 Autocorrelation time:                1579.0
 ==============================================================================
                            Augmented Dickey-Fuller Test
 ==============================================================================
 Test statistic for observable: -3.926148246630434
 P-value for observable: 0.001850619485090052
 The number of lags used: 46
 The number of observations used for the ADF regression: 35198
 Cutoff Metrics :
   1%: -3.430536 | The data is stationary with 99 % confidence
   5%: -2.861622 | The data is stationary with 95 % confidence
  10%: -2.566814 | The data is stationary with 90 % confidence
\end{lstlisting}

    The results dictionary contains the calculated MSE array (\texttt{'MSE'}), the equilibration time (\texttt{'t0'}), the average over the equilibrated part (\texttt{'average'}), the calculated uncertaunty (\texttt{'uncertainty'}), the equilibrated array (\texttt{'equilibrated'}), the autocorrelation time (\texttt{'ac\_time'}) and the number of uncorrelated samples (\texttt{'uncorr\_samples'}). These results can be directly utilized in automated workflows for data equilibration in simulations.

    \section*{Dependency on batch size}

\begin{figure}[!htb]
        \centering
        \includegraphics[width=.8\linewidth]{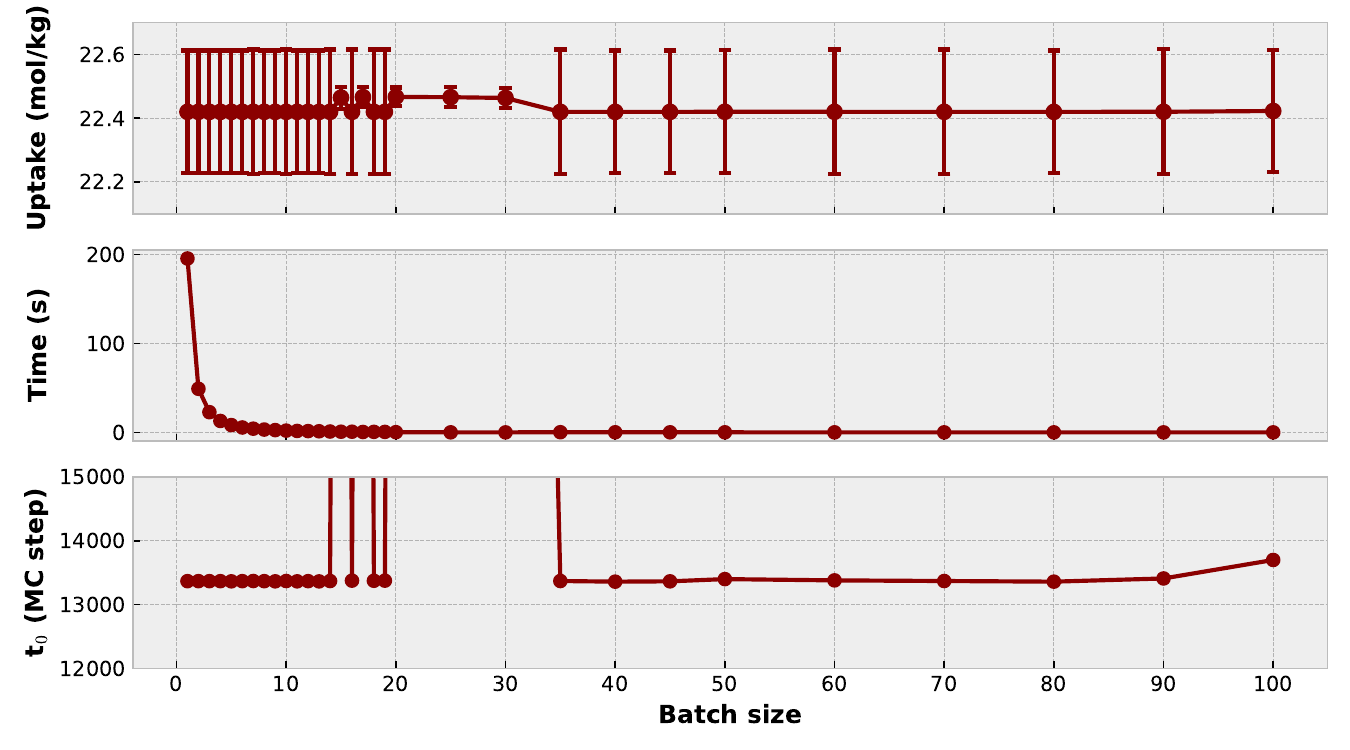}
        \caption{Dependency of the equilibrated uptake, execution time and truncation step on the batch size.}
        \label{fig:batch_size}
    \end{figure}

\begin{figure}[!htb]
        \centering
        \includegraphics[width=.8\linewidth]{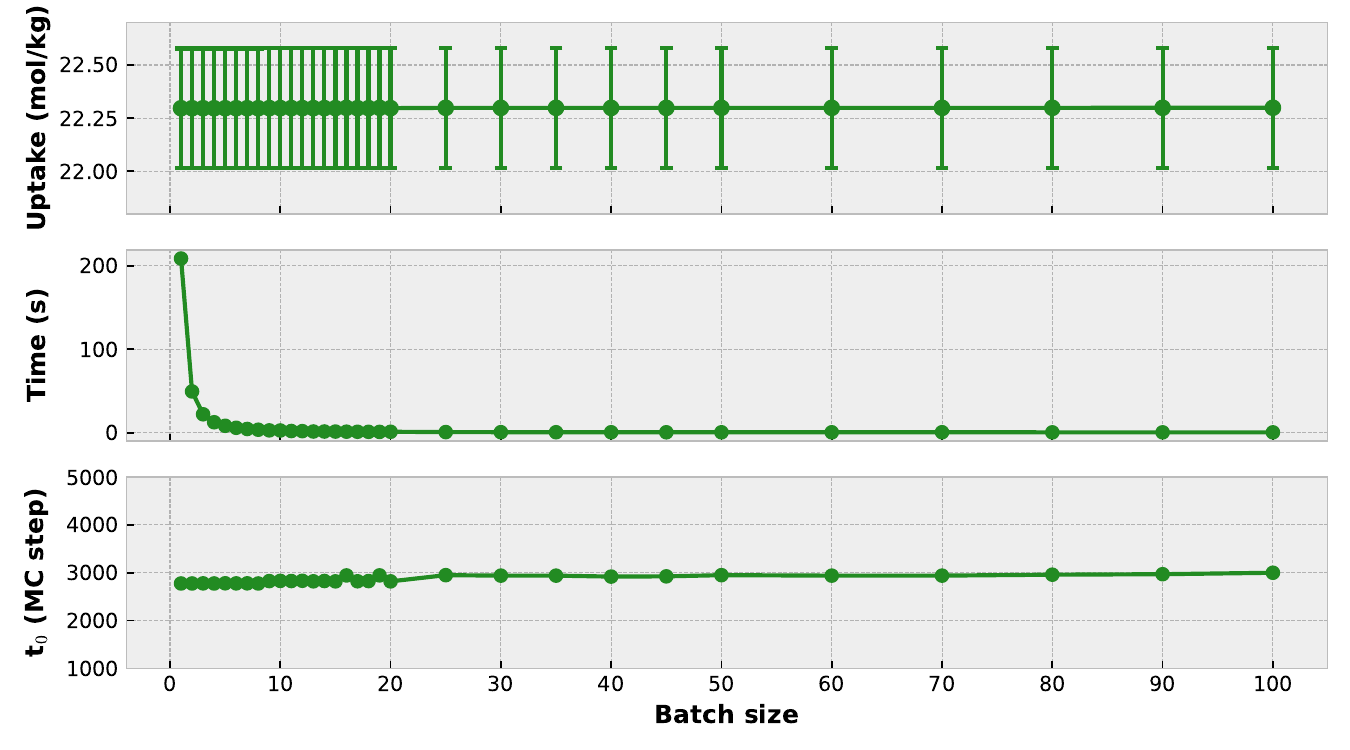}
        \caption{Dependency of the equilibrated uptake, execution time and truncation step on the batch size using the LLM version.}
        \label{fig:batch_size_LLM}
    \end{figure}

    The \autoref{fig:batch_size} displays the dependence of the mean balanced uptake value, pyMSER runtime, and truncation point obtained as a function of the selected batch size. It can be observed that the execution time decreases exponentially, reaching the scale of a few seconds with a batch size of approximately 5. In the region between 15 $<$ m $<$ 35, some instabilities in identifying the optimal truncation point of the time series can be observed. These instabilities can be completely resolved by using the \textbf{L}eft-most \textbf{L}ocal \textbf{M}inimum (LLM) version of the MSER method, as shown in \autoref{fig:batch_size_LLM}.

    \section*{Enthalpy of adsorption}

\begin{figure}[!htb]
        \centering
        \includegraphics[width=1\linewidth]{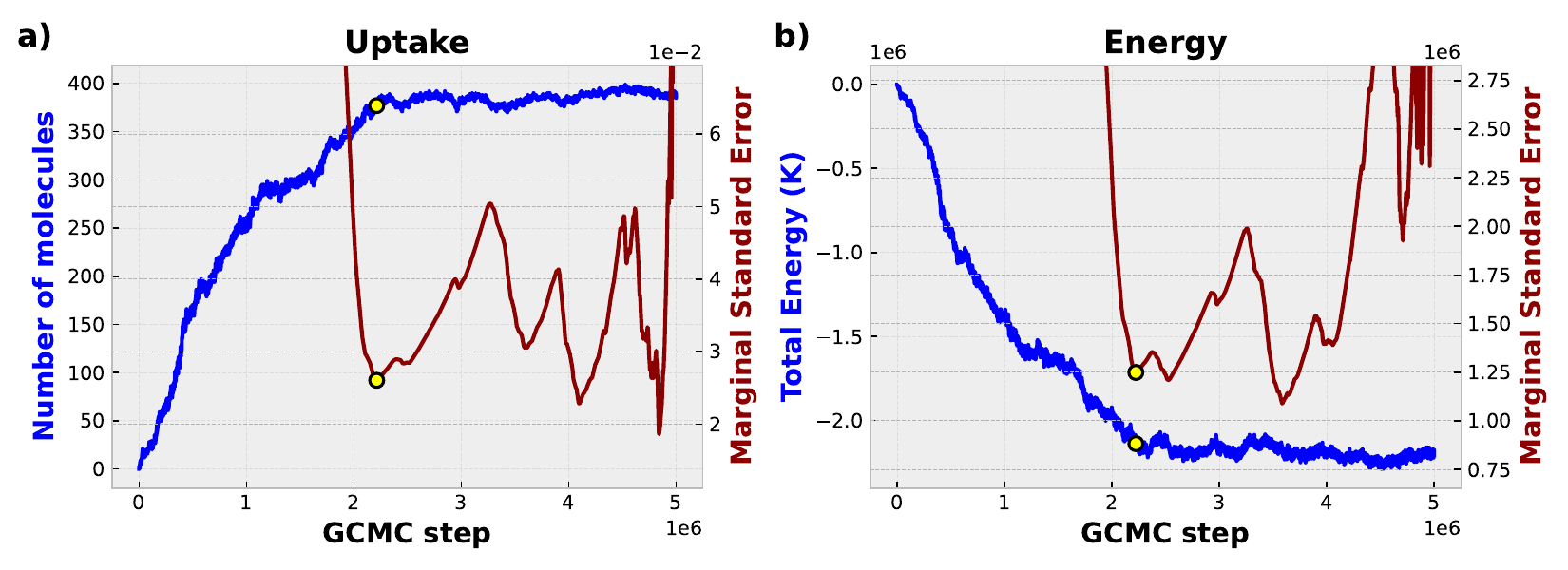}
        \caption{a) Uptake as function of MC step. b) Total energy as function of MC step for the adsorption of H\textsubscript{2}O on the HKUST-1 MOF at 198 K and 0.1 bar. }
        \label{fig:enthalpy}
    \end{figure}

    In \autoref{fig:enthalpy}(a), we display the H\textsubscript{2}O uptake overlaid to the marginal standard error as a function of GCMC steps. The leftmost local minimum of the MSE is highlighted, marking the start of the production stage. \autoref{fig:enthalpy}(b), in turn, displays the total energy overlaid to the MSE as a function of GCMC steps, where the truncation point is highlighted in a similar fashion. These results correspond to the simulation of H\textsubscript{2}O adsorption on the HKUST-1 MOF at 198 K and 0.1 bar.

\begin{figure}[!htb]
        \centering
        \includegraphics[width=0.5\linewidth]{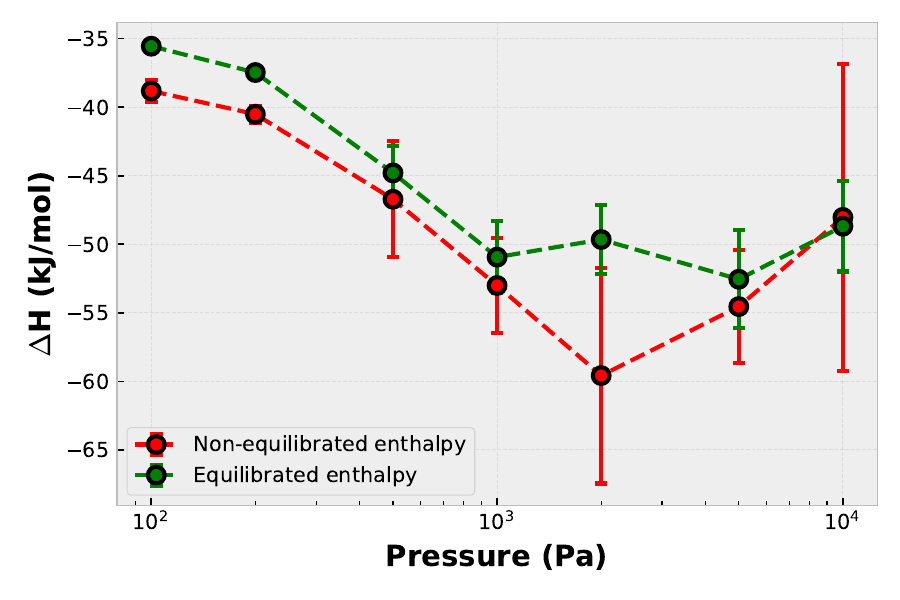}
        \caption{Comparison between equilibrated and non-equilibrated enthalpy of adsorption}
        \label{fig:enthalpy_comp}
    \end{figure}

    The impact of including the non-equilibrated part of the simulated time series in the enthalpy calculation is shown in \autoref{fig:enthalpy_comp}, for the same system mentioned above. We observe that, for most pressure conditions, the enthalpy values do not match even when the statistical uncertainties are taken into consideration.

    \section*{Test on different GCMC data}

\begin{figure}[!htb]
        \centering
        \includegraphics[width=1\linewidth]{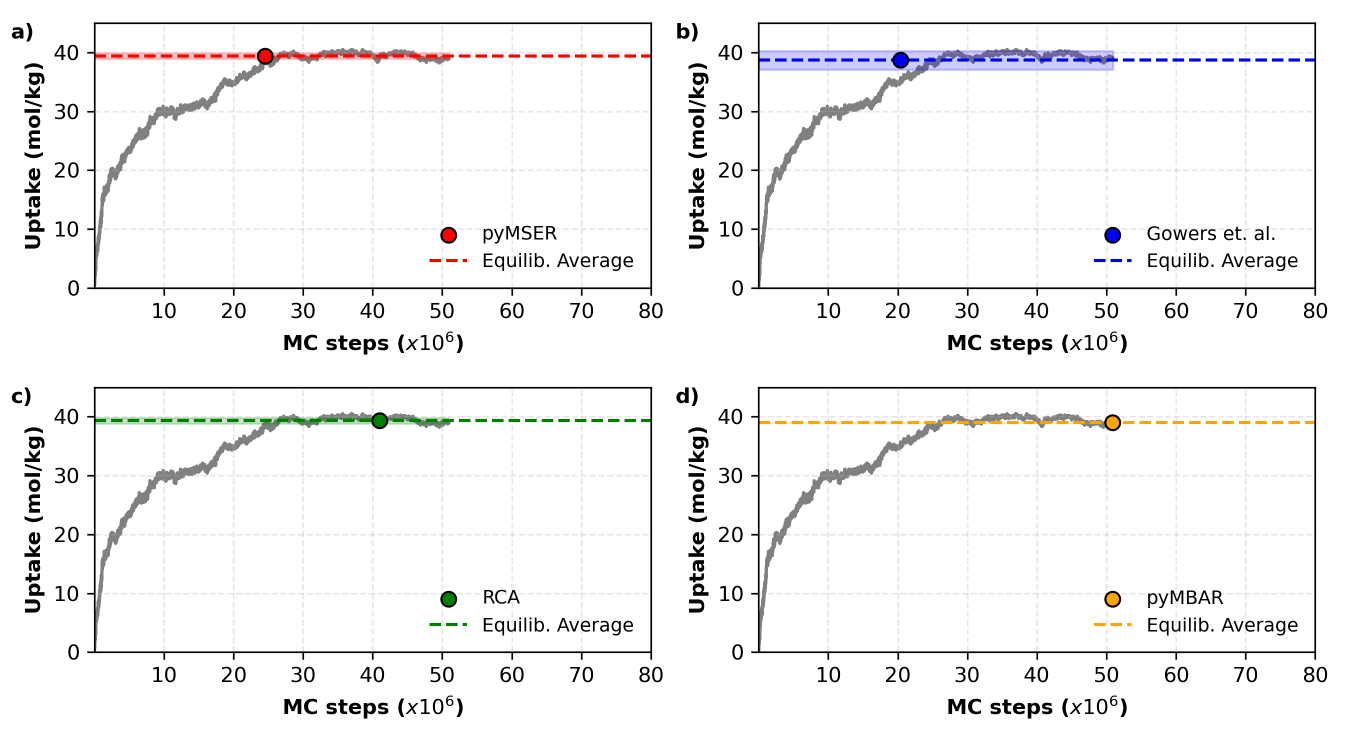}
        \caption{Truncation point detection with different approaches within a reduce dataset.}
        \label{fig:50m}
    \end{figure}

    In \autoref{fig:50m} we illustrate the results of the different methodologies on a slightly reduced dataset. It can been observed that the equilibration region now accounts for more than half of the data. As a result, all the methodologies used for analysis, except for pyMSER, have failed to detect the optimal equilibrium point for this dataset.
    
    In Figures S6-S10, we illustrate various GCMC simulation scenarios depicting the adsorption of CO\textsubscript{2} and H\textsubscript{2}O on widely recognized and extensively studied materials such as CuBTC MOF, IRMOF-1, and MgMOF-74. Notably, these simulations reveal a wide spectrum of equilibration patterns, ranging from fast equilibration with minimal standard deviation to very slow equilibration processes. 

\begin{figure}[!htb]
        \centering
        \includegraphics[width=1\linewidth]{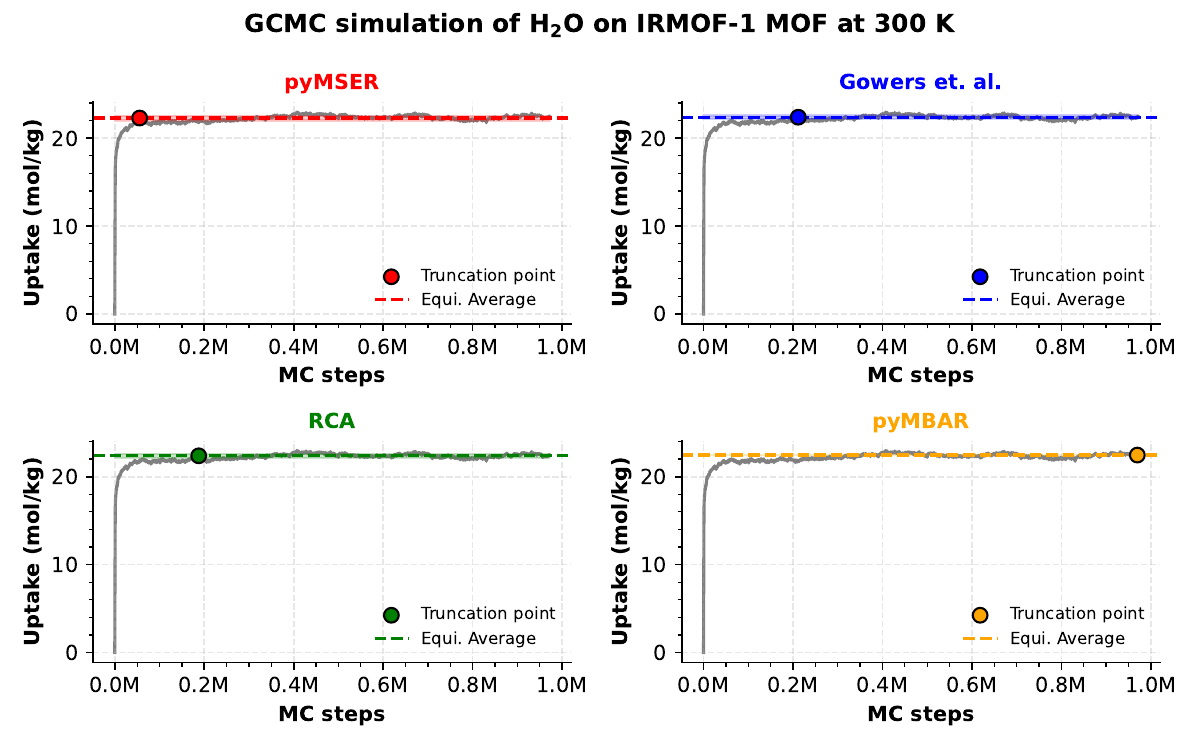}
        \caption{\textbf{GCMC simulation of H\textsubscript{2}O on IRMOF-1 at 300K} pyMSER, Gowers et. al. and RCA detect a reasonable truncation points on data that show rapid equilibration and low noise, while pyMBAR detects a very late truncation point.}
        \label{fig:GCMC_1}
    \end{figure}

\begin{figure}[!htb]
        \centering
        \includegraphics[width=1\linewidth]{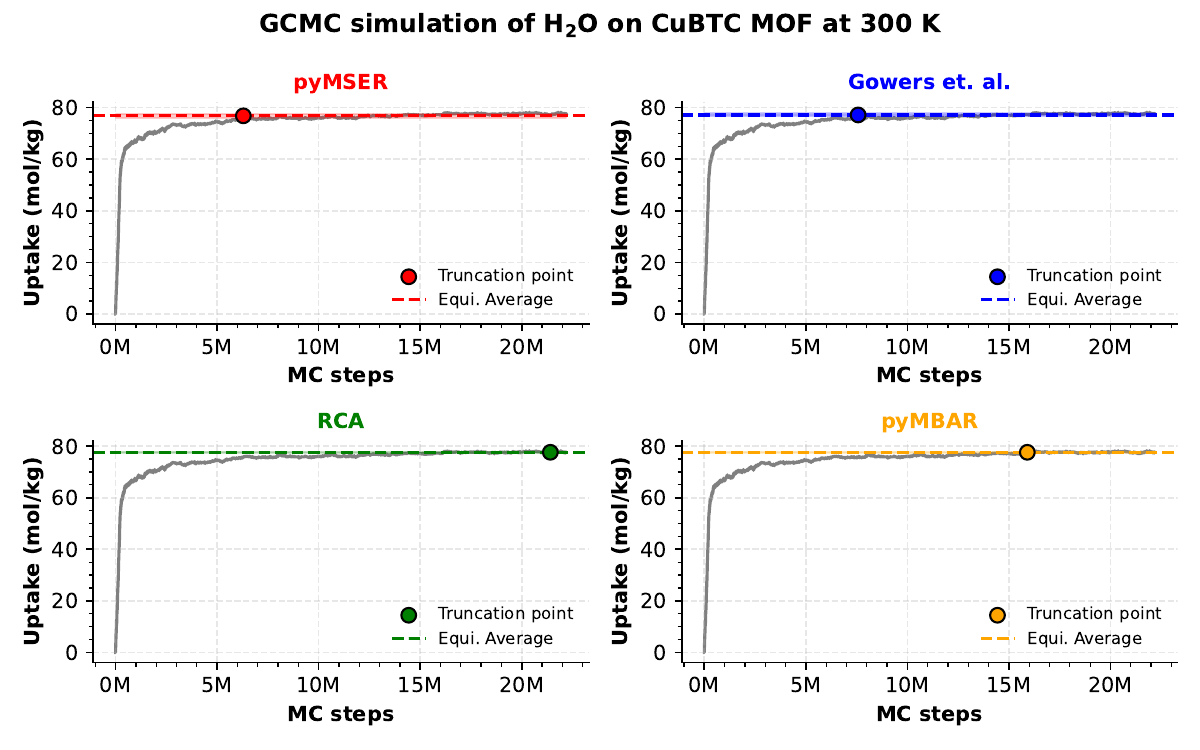}
        \caption{\textbf{GCMC simulation of H\textsubscript{2}O on CuBTC MOF at 300K} pyMSER, Gowers et. al. detect a reasonable truncation points on data that show intermediate equilibration and low noise while RCA and pyMBAR detects a very late truncation point.}
        \label{fig:GCMC_3}
    \end{figure}

    It is noteworthy that in fast equilibration and low standard deviation scenarios, as shown in the figures \autoref{fig:GCMC_1} and \autoref{fig:GCMC_3}, pyMBAR systematically detected quite late truncation points. The RCA method proved to be unstable, detecting a suitable truncation point on the rapid equilibration and low noise scenario and very late truncation points on the others.

\begin{figure}[!htb]
        \centering
        \includegraphics[width=1\linewidth]{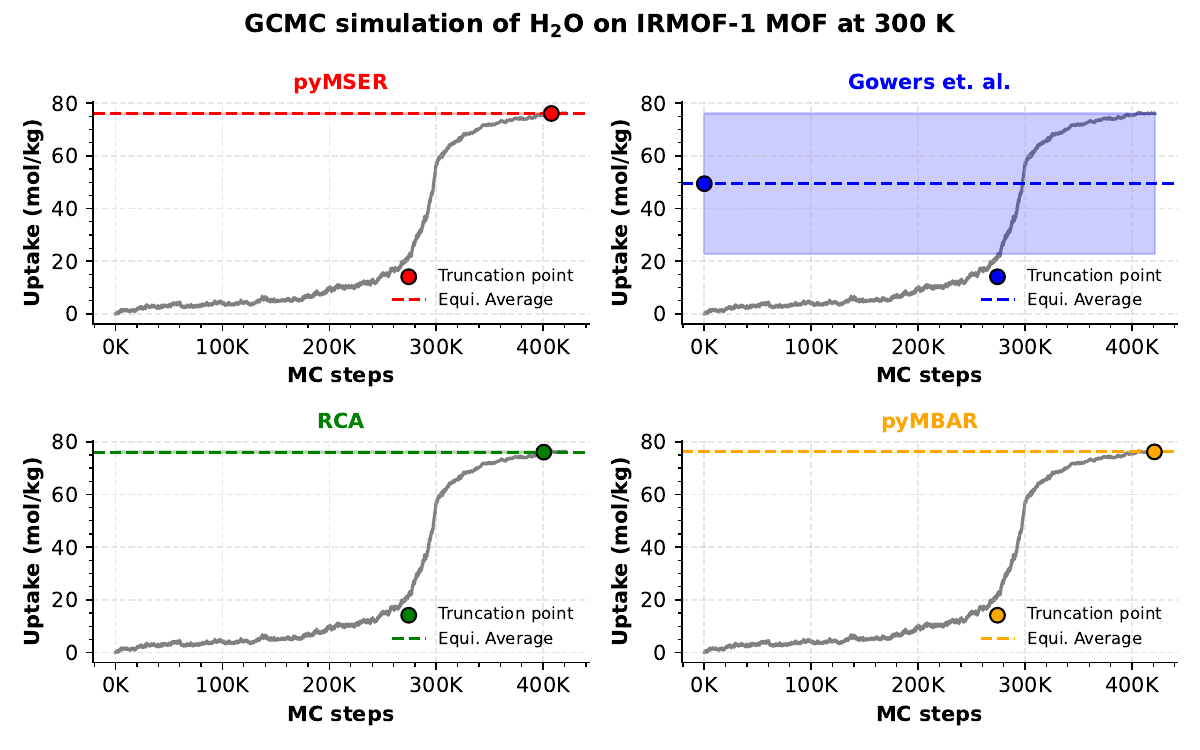}
        \caption{\textbf{GCMC simulation of H\textsubscript{2}O on IRMOF-1 at 300K} pyMSER, RCA and pyMBAR detect a reasonable truncation points on data that show slow equilibration and low noise while Gowers et. al. detects a very early truncation point.}
        \label{fig:GCMC_5}
    \end{figure}

    The method proposed by Gowers et. al. present good performance in most scenarios. However, it clearly fails when the basic condition of its application, which is that the second half of the data is in the equilibrated region, is not satisfied as illustrated in \autoref{fig:GCMC_5}.

    The methodology employed in pyMSER stood out as the sole approach capable of consistently identifying reasonable truncation points across all scenarios, irrespective of the equilibrium regime or standard deviation within the data set. This accomplishment can be attributed to pyMSER's global perspective on the data set, which aims to find an optimal balance between accuracy and sample size. In contrast, other methods follows a local approach, going in the reverse direction for the data examination, rendering them more susceptible to local fluctuations and thus less robust across different scenarios.

    \begin{figure}[!htb]
        \centering
        \includegraphics[width=1\linewidth]{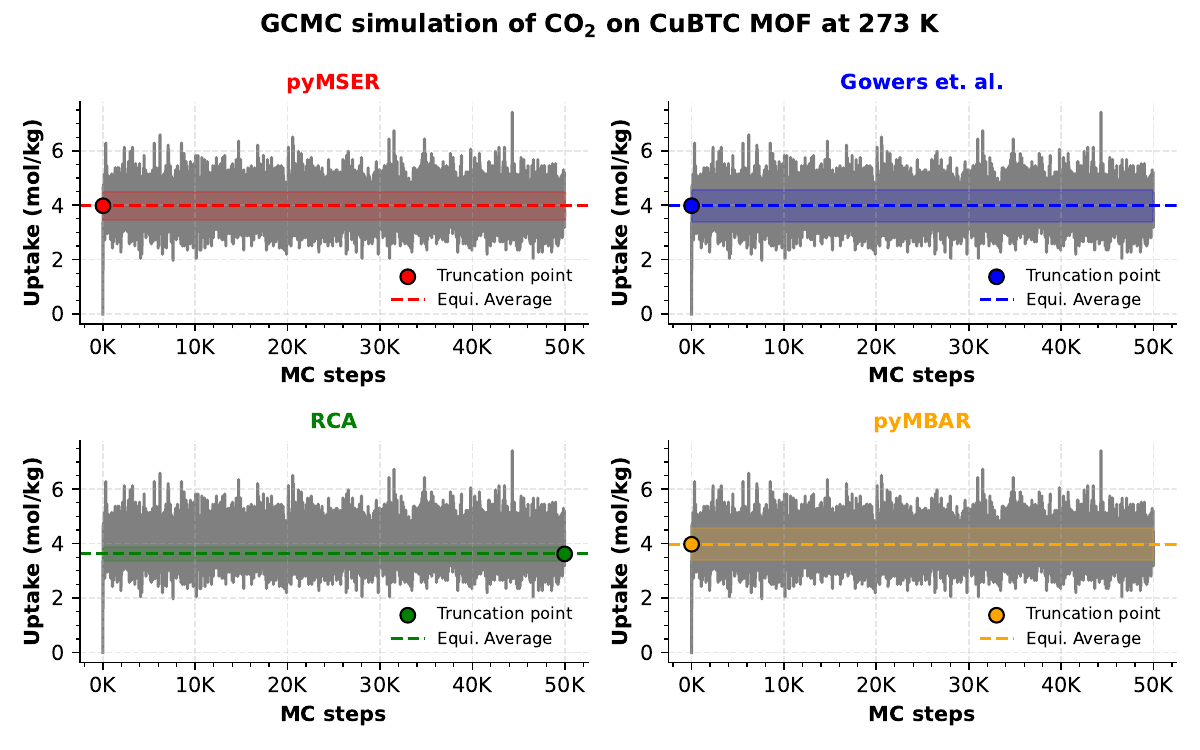}
        \caption{\textbf{GCMC simulation of CO\textsubscript{2} on CuBTC MOF at 273K} pyMSER, Gowers et. al. and pyMBAR detect a reasonable truncation points on data that show rapid equilibration and high noise while RCA detects a very late truncation point.}
        \label{fig:GCMC_2}
    \end{figure}

    \begin{figure}[!htb]
        \centering
        \includegraphics[width=1\linewidth]{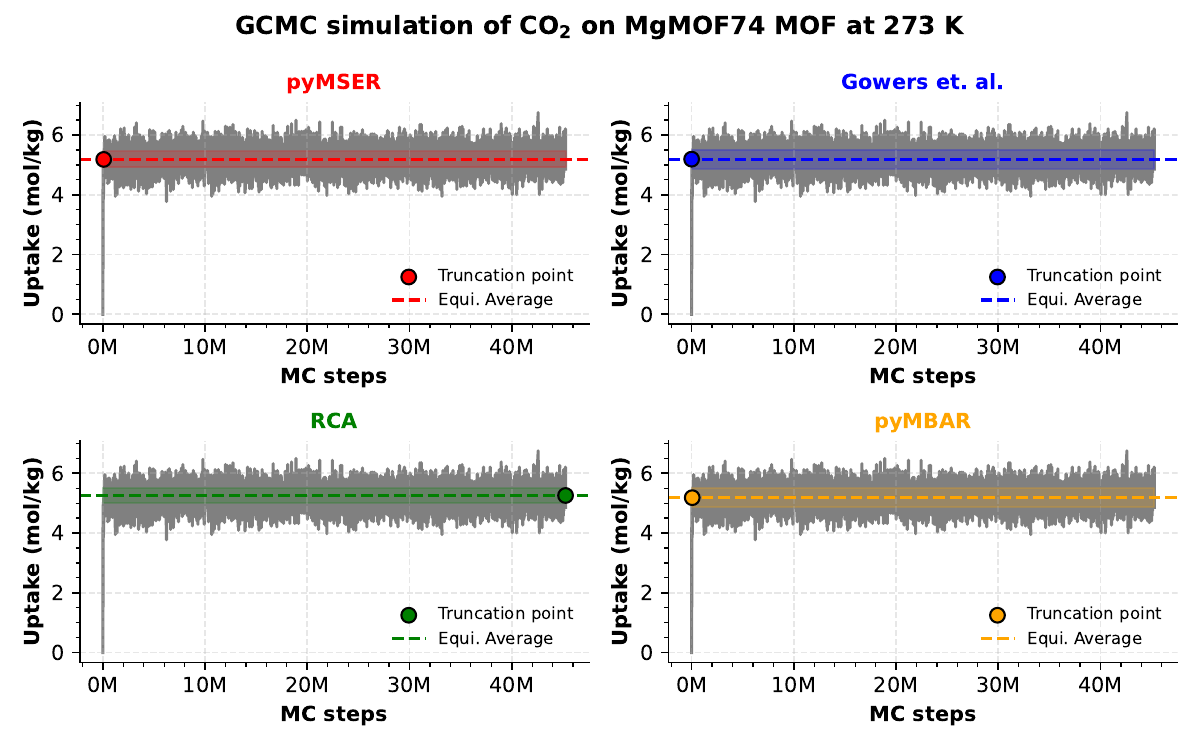}
        \caption{\textbf{GCMC simulation of CO\textsubscript{2} on MgMOF-74 at 273K} pyMSER, Gowers et. al. and pyMBAR detect a reasonable truncation points on data that show rapid equilibration and intermediate noise while RCA detects a very late truncation point.}
        \label{fig:GCMC_4}
    \end{figure}

\clearpage

\section*{Lennard-Jones parameters}
The Lennard-Jones parameters for the UFF force field used in the calculations for the framework atoms are shown in Table \ref{tab:uff}. The TraPPE and TIP5Ew parameters used for the gas molecules are present in Table \ref{tab:trappe}. The critical parameters used in the Peng-Robinson equation to calculate the fugacity are present in Table \ref{tab:critical}.

\begin{table}[!h]
\centering
    \begin{tabular}{@{}cccc@{}}
    \hline\hline 
    \textbf{Atom type} & $\sigma$ (\AA{}) & $\varepsilon$ (K) & \textbf{Charge} \\ \hline
    C\_CO2     & 2.80       & 27.0          & 0.700  \\
    O\_CO2     & 3.05       & 79.0          & -0.350 \\
    N\_N2      & 3.31       & 36.0          & -0.482 \\
    N\_com     & -          & -             & 0.964  \\
    Ow         & 3.097      & 89.63         & 0.964  \\
    Hw         & -          & -             & 0.241  \\ 
    Lw         & -          & -             & -0.241  \\ \hline\hline 
    \end{tabular}
    \caption{\label{tab:trappe} Lennard-Jones parameters for molecules.}
\end{table}

\begin{table}[!h]
    \centering
    \begin{tabular}{@{}cccc@{}}
    \hline\hline 
        \textbf{Gas} & \textbf{Critical temperature} (K) & \textbf{Critical Pressure} (Pa) & \textbf{Acentric factor} \\ \hline
        CO\textsubscript{2} & 304.1282                 & 7377300.0              & 0.22394         \\
        N\textsubscript{2}  & 126.192                  & 3395800.0              & 0.0372          \\
        H\textsubscript{2}O & 304.1282                 & 7377300.0              & 0.22394         \\ \hline\hline 
    \end{tabular}
     \caption{\label{tab:critical} Critical parameters for CO\textsubscript{2} and N\textsubscript{2}.}
\end{table}

\begin{table}[]
\centering
\begin{tabular}{ccc|ccc|ccc}
\hline\hline 
Atom & $\varepsilon$ (K) & $\sigma$ (\AA{}) & Atom & $\varepsilon$ (K) & $\sigma$ (\AA{}) & Atom & $\varepsilon$ (K) & $\sigma$ (\AA{}) \\ \hline 
B\_       & 47.804      & 3.582     & Ni\_      & 7.548       & 2.525     & Fr\_                      & 25.16                      & 4.366     \\
C\_       & 47.854      & 3.474     & Cu\_      & 2.516       & 3.114     & Ra\_                      & 203.293                    & 3.276     \\
H\_       & 7.649       & 2.847     & Zn\_      & 27.676      & 4.045     & La\_                      & 8.554                      & 3.138     \\
N\_       & 38.948      & 3.263     & In\_      & 276.76      & 4.09      & Ce\_                      & 6.542                      & 3.169     \\ 
O\_       & 48.156      & 3.034     & Sn\_      & 276.76      & 3.983     & Pr\_ & 5.032 & 3.213     \\ 
P\_       & 161.024     & 3.698     & Sb\_      & 276.76      & 3.876     & Nd\_                      & 5.032                      & 3.186     \\
S\_       & 173.101     & 3.591     & Te\_      & 286.824     & 3.769     & Pm\_                      & 4.529                      & 3.161     \\
F\_       & 36.482      & 3.094     & Rb\_      & 20.128      & 3.666     & Sm\_                      & 4.026                      & 3.137     \\
I\_       & 256.632     & 3.698     & Sr\_      & 118.252     & 3.244     & Eu\_                      & 4.026                      & 3.112     \\
K\_       & 17.612      & 3.397     & Zr\_      & 34.721      & 2.784     & Gd\_                      & 4.529                      & 3.001     \\
V\_       & 8.051       & 2.801     & Nb\_      & 29.689      & 2.82      & Tb\_                      & 3.522                      & 3.075     \\
W\_       & 33.714      & 2.735     & Mo\_      & 28.179      & 2.72      & Dy\_                      & 3.522                      & 3.055     \\
Y\_       & 36.23       & 2.981     & Tc\_      & 24.154      & 2.671     & Ho\_                      & 3.522                      & 3.038     \\
U\_       & 11.07       & 3.025     & Ru\_      & 28.179      & 2.64      & Er\_                      & 3.522                      & 3.022     \\
Li\_      & 12.58       & 2.184     & Rh\_      & 26.67       & 2.61      & Tm\_                      & 3.019                      & 3.006     \\
Be\_      & 42.772      & 2.446     & Pd\_      & 24.154      & 2.583     & Yb\_                      & 114.73                     & 2.99      \\
Al\_      & 155.992     & 3.912     & Ag\_      & 18.115      & 2.805     & Lu\_                      & 20.631                     & 3.243     \\
T\_       & 155.992     & 3.805     & Cd\_      & 114.73      & 2.538     & Ac\_                      & 16.606                     & 3.099     \\
Si\_      & 155.992     & 3.805     & Tl\_      & 342.176     & 3.873     & Th\_                      & 13.083                     & 3.026     \\
Cl\_      & 142.557     & 3.52      & Pb\_      & 333.622     & 3.829     & Pa\_                      & 11.07                      & 3.051     \\
Na\_      & 251.6       & 2.801     & Bi\_      & 260.658     & 3.894     & Np\_                      & 9.561                      & 3.051     \\
Mg\_      & 55.855      & 2.692     & Po\_      & 163.54      & 4.196     & Pu\_                      & 8.051                      & 3.051     \\
Ga\_      & 201.28      & 3.912     & At\_      & 142.909     & 4.233     & Am\_                      & 7.045                      & 3.013     \\
Ge\_      & 201.28      & 3.805     & Rn\_      & 124.794     & 4.246     & Cm\_                      & 6.542                      & 2.964     \\
As\_      & 206.312     & 3.698     & Cs\_      & 22.644      & 4.025     & Bk\_                      & 6.542                      & 2.975     \\
Se\_      & 216.376     & 3.591     & Ba\_      & 183.165     & 3.3       & Cf\_                      & 6.542                      & 2.952     \\
Br\_      & 186.184     & 3.52      & Hf\_      & 36.23       & 2.799     & Es\_                      & 6.038                      & 2.94      \\
Ca\_      & 25.16       & 3.094     & Ta\_      & 40.759      & 2.825     & Fm\_                      & 6.038                      & 2.928     \\
Sc\_      & 9.561       & 2.936     & Re\_      & 33.211      & 2.632     & Md\_                      & 5.535                      & 2.917     \\
Ti\_      & 8.554       & 2.829     & Os\_      & 18.618      & 2.78      & No\_                      & 5.535                      & 2.894     \\
Cr\_      & 7.548       & 2.694     & Ir\_      & 36.734      & 2.531     &                           &                            &           \\
Mn\_      & 6.542       & 2.638     & Pt\_      & 40.256      & 2.454     &                           &                            &           \\
Fe\_      & 27.676      & 4.045     & Au\_      & 19.625      & 2.934     &                           &                            &           \\
Co\_      & 7.045       & 2.559     & Hg\_      & 193.732     & 2.41      &                           &                            &       \\ \hline \hline 
\end{tabular}
\caption{\label{tab:uff} Lennard-Jones parameters for UFF force field.}
\end{table}

\section*{Alternatives to the Augmented Dickey-Fuller test}

    The Gelman-Rubin factor (GR) method provides a formula for an $R$ factor, which can be used to determine whether a simulation has reached convergence by comparing it to a user-defined criterion, $R^*$. In the version of the GR method presented by Anstine \textit{et al.}, the transient stage is excluded before performing the GR analysis. This exclusion is accomplished by fitting an exponential curve to the time series data and identifying the truncation point where the derivative falls below an arbitrary threshold, set at $10^{-4}$. Both the $R^*$ value and the derivative threshold were empirically determined through tests on GCMC data.

\begin{figure}[!htb]
        \centering
        \includegraphics[width=1\linewidth]{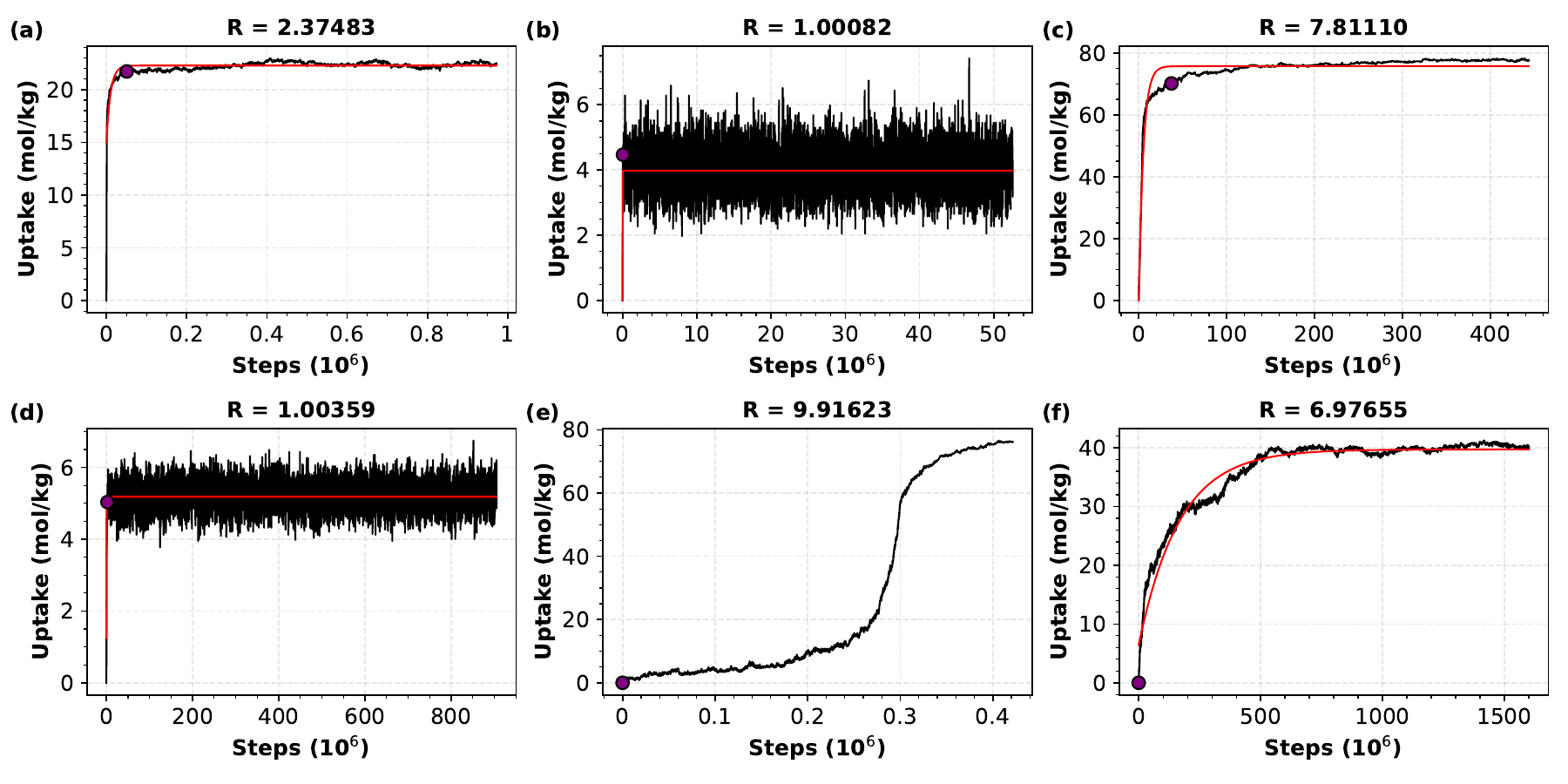}
        \caption{\textbf{Gelman-Rubin factor applied to several GCMC data.} Black lines represent the total uptake, red lines represent the fitted exponential curve, and the purple dots are the equilibration points detected by the Gelman-Rubin factor. The GCMC data corresponds to a) H\textsubscript{2}O on IRMOF-1 at 300K, b) CO\textsubscript{2} on CuBTC MOF at 273K, c) H\textsubscript{2}O on  CuBTC MOF at 300K, d) CO\textsubscript{2} on MgMOF74 at 273K, e) H\textsubscript{2}O on IRMOF-1 at 300K, and f) H\textsubscript{2}O on IRMOF-1 at 300K.}
        \label{fig:gelman_rubi}
    \end{figure}

    The application of the above method to GCMC example data is illustrated in \autoref{fig:gelman_rubi}. For cases with extremely rapid equilibrium, such as (b) and (d), the truncation point detection aligns with the pyMSER method, and the resulting $R$ value is close to 1, indicating correctly that the production data selected is equilibrated. 
    
    In the data shown in \autoref{fig:gelman_rubi}(a), the truncation point is correctly identified. However, the $R$ value is significantly higher than 1.5. This is unexpected, as both visual inspection and the Augmented Dickey-Fuller test suggest that this region is satisfactorily equilibrated. If we had used the standard criterion of $R^*=1.5$, we would have incorrectly classified this region as non-equilibrated.

\begin{figure}[!htb]
        \centering
        \includegraphics[width=1\linewidth]{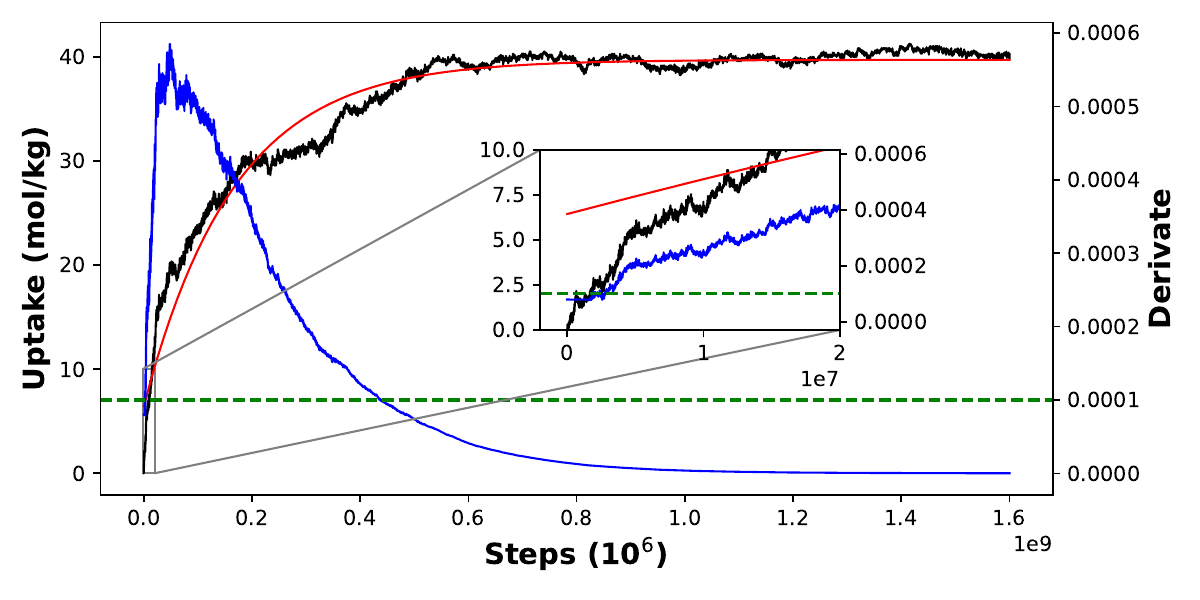}
        \caption{\textbf{Detection of the truncation point based on the fit of an exponential curve on the uptake data.} The threshold for the derivative curve leads to detecting an incorrect truncation point.}
        \label{fig:fit_exp_decay}
    \end{figure}

    Cases (c) and (f) presented incorrect values for the truncation point that were specifically caused by fitting an exponential curve to the time series data. In case (c), the data does not precisely follow an exponential model, causing the fitted curve to poorly represent points near the transient-production transition as shown in \autoref{fig:fit_exp_decay}. For case (f), on the start of the fitted curve the derivative is smaller than the imposed threshold of $10^{-4}$, causing this method to incorrectly select a premature truncation point. Thus, the calculated $R$ value is also incorrect and, consequently, any conclusions drawn from this analysis will also be incorrect.

    For case (e), the large $R$ value indicates that the curve did not reach equilibrium, and neither is it possible to fit an exponential curve to the data. This results in the detection of the truncation point to be extremely premature. 

    This, again, reinforces one of the value propositions of pyMSER: a parameter-free method that analyses the whole time series to minimise an error function. Even though the GR method has proven to be useful in many scenarios, the dependence on carefully-determined \textit{ad hoc} parameters makes it less suitable for an automated high-throughput screening use case. Although extensive tests can be conducted to optimally determine these parameters, one cannot guarantee that they will always work in a scenario were manual data inspection is practically impossible.

\end{document}